\documentclass{article}

\usepackage{arxiv}

\usepackage[utf8]{inputenc} 
\usepackage[T1]{fontenc}    

\usepackage[backend=biber,style=apa,autocite=inline]{biblatex}
\DeclareLanguageMapping{english}{english-apa}
\usepackage{hyperref}       
\usepackage{url}            
\usepackage{booktabs}       
\usepackage{amsmath}
\usepackage{amsfonts}       
\usepackage{nicefrac}       
\usepackage{microtype}      
\usepackage{xcolor}         

\usepackage{tabularx}
\usepackage{adjustbox}
\usepackage{pdflscape}
\usepackage{array}
\usepackage{tabularray} 

\usepackage{enumitem}
\usepackage{makecell}

\usepackage[table]{xcolor}

\newlist{tablelist}{itemize}{1}
\setlist[tablelist]{label=\textbullet, leftmargin=*, nosep, after=\vspace{-\baselineskip}, before=\vspace{-0.8\baselineskip}}

\usepackage{xltabular}
\newcommand{\defn}[1]{\par\vspace{1pt}%
  {\hangindent=1em \hangafter=0 \noindent\scriptsize\itshape #1\par}}
 
\usepackage{tikz}
\usetikzlibrary{shapes.geometric, arrows, arrows.meta, positioning, calc, shapes.geometric, shadows}
\usepackage{graphicx}

\AtEveryCitekey{\clearlist{language}}
\AtEveryBibitem{\clearlist{language}}

\AtEveryCitekey{\ifentrytype{article}{\clearfield{publisher}{}}{}}
\AtEveryCitekey{\ifentrytype{preprint}{\clearfield{publisher}{}}{}}
\AtEveryCitekey{%
  \ifboolexpr{
    test {\ifentrytype{book}}
    or
    test {\ifentrytype{article}}
  }{%
    \clearfield{month}%
    \clearfield{day}%
  }{}%
}

\AtEveryBibitem{\clearlist{address}}
\AtEveryCitekey{\clearlist{address}}
\AtEveryBibitem{\clearlist{location}}
\AtEveryCitekey{\clearlist{location}}

\DeclareSourcemap{
  \maps[datatype=bibtex]{
    \map{
      \step[fieldsource=doi, final]
      \step[fieldset=url, null]
    }
  }
}

\DeclareSourcemap{
  \maps[datatype=bibtex]{
    \map{
      \step[fieldsource=date, final]
      \step[fieldset=urldate, null]
    }
    \map{
      \step[fieldsource=year, final]
      \step[fieldset=urldate, null]
    }
  }
}

\DeclareSourcemap{
  \maps[datatype=bibtex]{
    \map{
      \step[fieldsource=title, match=\regexp{^([^,:\!\?!]+)}, final]
      \step[fieldset=shorttitle, origfieldval]
    }
  }
}

\title{Mapping U.S. Federal AI Governance Against Sector Vulnerability}

\author{%
  Ho Ting (Bosco) Hung\thanks{Corresponding author: boscohht@mit.edu} \\
  MARS 
\And
  Angelica Chowdhury \\
  MARS
\And
  James Teague \\
  MARS
\And
  Simon Mylius \\
  MIT FutureTech
\And
 Spencer Michaels \\
 Cambridge Boston Alignment Initiative
\And
  Peter Slattery \\
  MIT FutureTech
\And
  Alexander Saeri \\
  MIT FutureTech
  \And
  Neil Thompson \\
  MIT FutureTech
}

\begin{document}
\maketitle

\begin{abstract}
Artificial intelligence (AI) poses different levels of risk across sectors, but are these differences reflected in U.S. federal AI governance? To help answer this question, we assess 684 federal AI governance documents for their coverage of 14 sectors and 24 AI risks. We measure coverage as breadth (i.e., how frequently the risk or sector is addressed across documents) and depth (i.e., how substantively the risk or sector is discussed). We then compare sector coverage patterns for each of the 24 risks with vulnerability assessments from a Delphi study of 272 experts. Our analysis finds substantial variation in coverage: AI risks related to robustness, system security, and governance receive more attention than socioeconomic, environmental, and emerging risks, including multi-agent risks. Public administration, national security, information, and scientific services receive comparatively high levels of coverage relative to other sectors, such as finance and healthcare, which experts rate as highly vulnerable to AI risks. By mapping current coverage and identifying where it differs from expert assessments of vulnerability, we surface potential AI governance gaps which may help inform AI risk-related decisions across government and industry. 
\end{abstract}

\newpage
\tableofcontents
\listoffigures
\listoftables

\newpage
\section*{Plain Language Summary}
\addcontentsline{toc}{section}{Plain Language Summary}
This paper explores which sectors and AI risks are covered in 684 U.S. federal governance documents and contrasts this coverage against expert assessments of each sector's risk vulnerability.

\subsection*{What We Did}
We analyzed 684 U.S. federal AI governance documents with major activity between January 2020 and January 20, 2026. The documents were drawn from the AGORA dataset and classified according to the 14 sectors and 24 AI risk subdomains they addressed. For each sector and risk, we measured both how often it was mentioned and how much detail the documents provided, including whether they described specific governance measures. We then compared these coverage patterns with sector vulnerability ratings from a three-round survey of 272 experts in a Delphi study. 

\subsection*{What We Found}

Our findings reveal substantial differences in how often sectors and AI risks are addressed in U.S. federal AI governance documents and in how much detail they are discussed. In some cases, the distribution of documentary attention differs from expert assessments of sector vulnerability.

\begin{itemize}
    \item \textbf{Attention is unevenly distributed across sectors:}
    Public administration, national security, information, and scientific services receive comparatively high levels of coverage. By comparison, finance and insurance, healthcare, and social assistance receive less AI-specific documentary attention, despite receiving relatively high vulnerability ratings from experts.

    \item \textbf{Most risks and sectors receive limited detailed treatment:}
    Most documents that mention a sector or risk subdomain devote no more than a few sentences to it. For nearly every risk subdomain, fewer than half of the documents mentioning a particular subdomain meet the study's threshold for good coverage. This suggests that the documents analyzed generally do not provide detailed and explicit governance measures.

    \item \textbf{Socioeconomic, environmental, and emerging risks receive comparatively little attention:}
    Risks related to system security, robustness, and governance are addressed more frequently than most socioeconomic, environmental, and emerging risks. Risks such as environmental harm, the devaluation of human effort, and multi-agent interactions are rarely discussed and seldom receive detailed treatment when they are mentioned.
\end{itemize}

These findings describe patterns within AI-specific federal governance documents. By revealing these differences, this paper offers insights to inform AI policy research and also high-level decision-making in various sectors, especially public administration. However, the findings do not necessarily indicate that the identified sectors or risks are inadequately governed, since relevant protections may also be provided through generally applicable laws, sector-specific regulation, state and local measures, private standards, or other governance mechanisms outside those examined.

\subsection*{Why This Matters}

The findings provide a starting point for policymakers and researchers seeking to understand how federal attention is distributed across AI risks and sectors. They highlight areas where further investigation may be useful, including examining whether existing laws and other governance mechanisms adequately address risks.

\newpage
\section{Introduction}
There is a growing concern that governance is failing to keep pace with developments in artificial intelligence (AI) \autocite{wirtzGovernanceArtificialIntelligence2022, bengio_international_2025}. One contributing challenge is that regulators and key actors often lack a shared understanding of AI risks, which leads to variations in governance approaches across institutions and jurisdictions, and thus disparities in potential oversight \autocite{zaidan_ai_2024, saeriMappingAIRisk2025, monteiro_wheel_2025, taeihaghGovernanceGenerativeAI2025, perboli_navigating_2025, slattery_ai_2026}. These issues are particularly relevant in the U.S., where AI governance measures are applied in many forms, including executive actions, legislation, regulations, agency guidance, standards, and other policy instruments \autocite{davtyan_us_2025}. 

One way to identify where governance is comparatively concentrated or lacking is to synthesize and evaluate collective coverage against some established expectation \autocite{myliusMappingAIGovernance2025}. To support this type of analysis, the MIT AI Risk Initiative developed the AI Governance Mapping Project, which classifies more than 1,000 AI-related governance documents according to the risks, sectors, actors, and a few other dimensions they address \autocite{myliusMappingAIGovernance2025, mylius_mapping_2026}. The initiative also conducted a 2025 AI Risk Prioritization Delphi study of 272 experts that assessed the relative severity of risks and the vulnerability of different sectors to each risk \autocite{saeri_prioritization_2026}. 

Building on these two projects, this paper asks: \textbf{How is risk coverage distributed across sectors and risk subdomains in U.S. federal AI governance documents, and how does this distribution compare with expert-rated sector vulnerability?} To answer this, we examine 684 federal AI governance documents with major activity between January 2020 and January 20, 2026. For each sector and risk subdomain, we measure both the breadth of coverage based on how frequently these two dimensions are addressed across documents, and the depth of coverage, based on how substantively these two dimensions are discussed. We then compare these patterns with expert assessments of sector vulnerability.

This paper makes two main contributions. First, it provides a systematic descriptive account of how AI-specific federal governance documents are distributed across sectors and risk subdomains. This complements existing work on the development and structure of U.S. AI governance, which has largely focused on policy trajectories, institutional arrangements, and governance philosophies \autocite{mallinson_artificial_2025, davtyan_us_2025, kalnina_tangled_nodate}. Second, it compares patterns of documentary coverage with expert-rated vulnerability to identify areas where they diverge. These comparisons are intended as a diagnostic tool for identifying sectors and risks that warrant further investigation.

The analysis reveals substantial variation in both the breadth and depth of coverage. Public administration, national security, information, and scientific services receive comparatively high levels of coverage, while finance and healthcare receive less AI-specific documentary coverage despite relatively high expert-rated vulnerability. Coverage also varies across risk categories: system security, robustness, and governance-related risks are addressed more frequently than most socioeconomic, environmental, and emerging risks. In addition, most documents that mention a sector or risk subdomain provide only limited detail under the study's coding rubric. Taken together, these findings identify differences between patterns of coverage in federal AI governance documents and expert assessments of sector vulnerability. 

This paper proceeds as follows. Section \ref{sec:methodology} describes the methodology structuring the analysis and contextualises the findings. Next, Section \ref{sec:finding} maps coverage across risks and sectors to evaluate it against expert assessments of vulnerability. Then, Section \ref{sec:discussion} examines the implications of the findings. Finally, Section \ref{sec:conclusion} provides the conclusion and outlines relevant future research avenues.

\newpage
\section{Methodology}
\label{sec:methodology}
\subsection{Data Sources}

\subsubsection{MIT AI Governance Mapping project}
\urldef{\agorascopeurl}\url{https://eto.tech/dataset-docs/agora-dataset/#scope}
This paper leverages the results of the MIT AI Risk Initiative's AI Governance Mapping project, which categorized 1000+ AI-relevant laws, standards, and other documents from the Center for Security and Emerging Technology’s (CSET) \href{https://eto.tech/dataset-docs/agora-dataset/}{AI Governance and Regulatory Archive (AGORA)} \autocite{arnoldIntroducingAIGovernance2024, arnoldAGORADataset2026}.\footnote{For more details on what was deemed `AI-relevant', see: \agorascopeurl.} AGORA only covers documents that directly and substantively address the development, deployment, or use of AI, including cases where general laws are tailored for AI. However, it typically excludes generally applicable laws with implications for AI. AGORA covers most enacted and proposed federal laws and regulations since 2020, the majority of executive orders since 2020, and other agency documents of interest.

The Governance Mapping project classified the AGORA documents by assigning coverage scores of 1-3 based on how extensively each document discussed or `covered' each of the 24 risk subdomains from the MIT AI Risk Domain Taxonomy \autocite{slattery_ai_2026}, 14 sectors based on the North American Industry Classification System, and other dimensions.\footnote{Before analyzing the full AGORA dataset, the Governance Mapping project compared how different large language models (LLMs) and expert human reviewers classified six AI-related governance documents from several U.S. jurisdictions. In doing so, it assessed the cost-effectiveness and inter-rater reliability of different approaches, as well as whether humans and LLMs generally agreed with one another. Claude Sonnet 4.5 was identified as the best-performing model amongst those tested. This model was then used to obtain the full classification results.} To be classified as having good coverage, a governance document needs to have clear and specific references to the dimension with at least multiple sentences of description. Table \ref{tab:coverage_rubric} shows the details of the rubric, which is the cornerstone of the breadth and depth of legislative coverage discussion in Section \ref{sec:finding}.

\begin{table}[htbp]
\centering
\caption{Coverage Score Rubric for Risk Subdomains and Sectors}
\label{tab:coverage_rubric}
\small
\begin{tabularx}{\textwidth}{@{} l X X @{}}
\toprule
\textbf{Score} & \textbf{Risk Subdomain Coverage} & \textbf{Sector Coverage} \\
\midrule
\textbf{1 (None)} & \textbf{No Coverage:} Risk subdomain is not mentioned at all. & \textbf{No Mention:} The sector is not mentioned or referenced in any way. \\
\addlinespace
\textbf{2 (Minimal)} & \textbf{Minimal Coverage:} Brief mention (no more than a few sentences). No specific mitigations or governance measures are described. & \textbf{Minimal Coverage:} The sector is mentioned, but with little elaboration on specific AI governance (no more than a few sentences). \\
\addlinespace
\textbf{3 (Good)} & \textbf{Good Coverage:} Comprehensive governance measures or mitigations with clear procedures explicitly described (typically $\geq$ 1 paragraph). & \textbf{Good Coverage:} Explicit, clear description of governance measures, controls, or obligations (typically several dedicated sentences). \\
\bottomrule
\end{tabularx}

\vspace{1em}

\begin{quote}
\footnotesize
\textbf{Scoring Note on the Mention vs. Coverage Distinction:} \\
A subdomain is only marked as `covered' (Score 2 or 3) if the document addresses the \textit{risk} itself. For example, a document describing general governance structures does \textbf{not} count as coverage for `6.5 Governance Failure' unless it explicitly describes how those structures might fail or become ineffective.
\end{quote}
\end{table}

\subsubsection{MIT AI Risk Prioritization Delphi Study}
As a separate workstream to the AI Governance Mapping project, the MIT AI Risk Initiative conducted the 2025 AI Risk Prioritization Delphi Study. Using a three-round Delphi methodology, the study surveyed 272 experts in late 2025 across the same 24 categories of AI risks. Experts judged AI risk severity and the probability that they would materialize, sector (e.g., finance, information) and actor (e.g., developer, deployer) vulnerability to risk, actor responsibility to address risk, and overall concern \autocite{saeri_prioritization_2026}.\footnote{The Delphi method is a decision-making technique that gathers insights from a structured panel of experts through multiple, anonymous, iterative rounds. This helps prevent authority or reputation from dominating the process, while allowing the correction of individual bias. It also helps reach expert consensus beyond simple percentage agreement to enhance confidence in an organization’s risk management approach.} Respondents rated each sector’s vulnerability using a five-point scale (1=`Not at all vulnerable' to 5=`Extremely vulnerable'). This paper compares the results of the Delphi study against findings from the Governance Mapping project on AI risk and sector coverage. The Delphi study’s findings are used to contextualize the implications of uneven legislative coverage by showing how possible gaps in U.S. federal governance contrast with expert concern about AI risk for those sectors.

\subsection{Analysis}
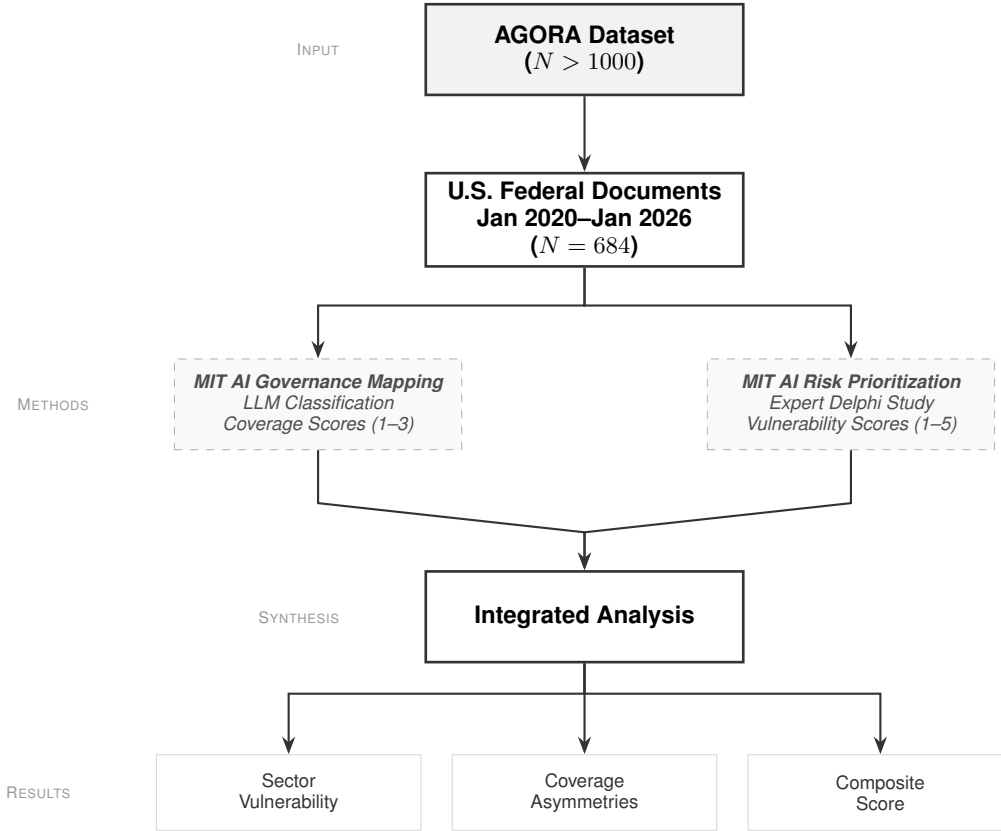
\begin{figure}[!htbp]
    \centering
    \begin{tikzpicture}[
        node distance = 1.0cm and 0.8cm,
        mainBox/.style={
            rectangle, 
            draw=black!80, 
            fill=white, 
            line width=1pt, 
            minimum width=4.2cm, 
            minimum height=1.2cm, 
            align=center, 
            font=\sffamily\bfseries\small
        },
        methodBox/.style={
            rectangle, 
            draw=black!30, 
            fill=gray!5, 
            dashed,
            minimum width=3.8cm, 
            minimum height=1.2cm,
            align=center, 
            font=\sffamily\scriptsize\itshape\color{black!70}
        },
        outputBox/.style={
            rectangle, 
            draw=black!15, 
            fill=white, 
            minimum width=3.5cm, 
            minimum height=1cm,
            align=center, 
            font=\sffamily\scriptsize\color{black!80}
        },
        arrowLine/.style={
            -{Stealth[scale=1.0]}, 
            line width=0.8pt, 
            color=black!80
        },
        thinLine/.style={
            black!30, 
            thin
        }
    ]

    \node[mainBox, fill=black!5] (agora) {\textbf{AGORA Dataset} \\ ($N>1000$)};
    \node[mainBox, below=of agora] (filter) {\textbf{U.S. Federal Documents} \\ Jan 2020--Jan 2026 \\ ($N=684$)};
    \node[mainBox, below=4cm of filter] (integrated) {\textbf{Integrated Analysis}};

    \node[methodBox, below left=1.2cm and -0.5cm of filter] (mapping) {
        \textbf{MIT AI Governance Mapping} \\ 
        LLM Classification \\ 
        Coverage Scores (1--3)
    };
    
    \node[methodBox, below right=1.2cm and -0.5cm of filter] (delphi) {
        \textbf{MIT AI Risk Prioritization} \\ 
        Expert Delphi Study \\ 
        Vulnerability Scores (1--5)
    };

    \node[outputBox, below=1.2cm of integrated] (output2) {Coverage \\ Asymmetries};
    \node[outputBox, left=0.4cm of output2] (output1) {Sector \\ Vulnerability};
    \node[outputBox, right=0.4cm of output2] (output3) {Composite \\ Score};

    
    \draw[arrowLine] (agora) -- (filter);
    
    \draw[arrowLine] (filter.south) -- ++(0,-0.5) -| (mapping.north);
    \draw[arrowLine] (filter.south) -- ++(0,-0.5) -| (delphi.north);
    
    \draw[arrowLine] (mapping.south) |- ++(0,-0.7) -- ($(integrated.north) + (0,0.5)$) -- (integrated.north);
    \draw[arrowLine] (delphi.south) |- ++(0,-0.7) -- ($(integrated.north) + (0,0.5)$) -- (integrated.north);

    \draw[arrowLine] (integrated.south) -- ++(0,-0.4) -| (output1.north);
    \draw[arrowLine] (integrated.south) -- (output2.north);
    \draw[arrowLine] (integrated.south) -- ++(0,-0.4) -| (output3.north);

    \node[font=\sffamily\scshape\tiny, color=black!40, left=1cm of agora] {Input};
    \node[font=\sffamily\scshape\tiny, color=black!40, left=1cm of mapping] {Methods};
    \node[font=\sffamily\scshape\tiny, color=black!40, left=1cm of integrated] {Synthesis};
    \node[font=\sffamily\scshape\tiny, color=black!40, left=1cm of output1] {Results};

    \end{tikzpicture}
    \caption{Methodological Framework}
    \label{fig:method_flow}
\end{figure}

Using the data sources outlined above, this paper performs the analysis below on the U.S. federal AI governance landscape, as summarized in Figure \ref{fig:method_flow}. We only include documents from AGORA that apply to U.S. jurisdictions and are proposed by federal authorities. We further limit the included documents to those with their last major activity occurring between January 2020 and January 20, 2026 (e.g., proposal, enactment). Together, these filters yield a corpus confined to a single legal and institutional setting and to a period of federal AI policy activity, so as to support a consistent read of patterns across documents. Since the corpus includes both enacted and proposed documents, the `coverage' analysis performed in this paper primarily measures documentary attention.

\subsubsection{Breadth and Depth of Coverage}
We measure coverage along two dimensions: breadth and depth. Breadth is the share of U.S. federal AI governance documents ($N=684$) that mention a given item---each of the 14 sectors, or each of the 24 risk subdomains in 7 risk domains. We define depth as the proportion of those mentioning documents that reach `good coverage’ of the particular risk subdomain or sector of interest, i.e., how substantively it is addressed. Comparing the two shows the varying levels of detail that are brought to a risk or sector, and how often it is discussed at all. 

\subsubsection{Comparing Coverage Against Sector Vulnerability}
These findings are further triangulated with expert analysis of sector vulnerability to examine the implications of coverage variations and identify potential divergences across various sectors. We summarize the misalignment between a sector's expert-rated vulnerability and its governance coverage using a single score per sector. We compute this in two ways, namely (1) an attention divergence, which compares vulnerability against coverage breadth alone, and (2) an integrated governance divergence, which folds in the depth of that coverage. 

Note that the composite scores are intended only to capture a relative mismatch between expert-perceived risk and federal documentary attention, rather than a demonstrated absence or failure of regulation. Since high documentary coverage or depth reflects only the volume and extent of formal policy discussion, a high divergence score does not automatically imply insufficient oversight, nor does extensive coverage guarantee effective governance in practice. Also, since expert-rated vulnerability and documentary share are not commensurable quantities, these composite scores should be read as a relative ranking of sectors, not as absolute estimates of the size of any governance gap or requirements of universal coverage.

\paragraph{Composite Attention Divergence Score}
Beyond a visual inspection of patterns, we compute a \textbf{composite attention divergence ($AD$)} score to quantify the coverage divergence across sectors by focusing on the sector share of mentions relative to expert-rated vulnerability. We first transform the raw vulnerability scores into a normalized score $v_{norm, i,j} = \frac{v_{score, i,j} - 1}{4} \in [0,1]$ for each subdomain $i$ and sector $j$ (denominator is 4 since vulnerability score ranges from 1 to 5). After standardizing sector vulnerability, we define the attention divergence for a specific subdomain-sector pair as the difference between the normalized vulnerability score and the sector share of coverage (defined below), $AD_{i,j} = v_{norm,i,j} - SectorShare_{i,j}$, so that a positive value marks under-attention (vulnerability exceeding coverage) and a negative value the reverse. We define sector share as the ratio of the joint breadth of the sector-subdomain pair ($B_{i,j}$) to the global breadth of the risk subdomain ($B_i$): 
$$SectorShare_{i,j} = \frac{B_{i,j}}{B_i} = \frac{\left( \frac{\sum \text{Documents Mentioning } (i \cap j)}{N} \right)}{\left( \frac{\sum \text{Documents Mentioning } i}{N} \right)}$$ 
Since both breadths are computed over the same $N=684$ documents, the normalization cancels and sector share reduces to the proportion of documents mentioning subdomain $i$ that also mention sector $j$. The resulting divergence indicates whether a sector’s share of the federal response to a risk is disproportionately low relative to its expert-rated vulnerability. Since sectors in general are assigned medium to high vulnerability ratings (typically $v_{score} > 2.5$) and reaching 100\% documentary coverage across the entire corpus is structurally unlikely, divergence is positive for most pairs by construction. However, this does not imply that a $SectorShare$ of 1.0 is the expected normative baseline because the breadth of coverage tends to be low (often $<50\%$) in practice, as legislative and policy documents are frequently narrow in scope or highly specific to certain sectors or subdomains.

To produce a single representative metric for each sector, we aggregate the $I=24$ individual subdomain divergences using the Root Mean Square (RMS). Because individual divergences are squared before averaging, the RMS weights large divergences more heavily than an arithmetic mean would. For two sectors with a comparable average divergence, the sector whose divergence is concentrated in a few subdomains receives the higher score. The final sector composite attention divergence score is calculated as follows:
$$CompositeAD_{j} = \sqrt{\frac{1}{I} \sum_{i=1}^{I} AD_{i,j}^2} = \sqrt{\frac{1}{I} \sum_{i=1}^{I} (v_{norm, i,j} - SectorShare_{i,j})^2}$$

\paragraph{Composite Integrated Governance Divergence Score}
While the attention divergence identifies the distribution of risk coverage across federal documents, a more comprehensive evaluation must also account for the depth of discussion within them. We therefore introduce the \textbf{integrated governance divergence ($IGD$)} score, which measures each (${i,j}$) pair's normalized vulnerability against a coverage value that combines sector share with the adjusted depth $D'_{i,j}$ (defined below) through a geometric mean: $IGD_{i,j}=v_{norm, i,j} - \sqrt{SectorShare_{i,j} \cdot D'_{i,j}}$, for each risk subdomain $i$ within sector $j$. The geometric mean, rather than an arithmetic average, penalises pairs where a sector covers a large share of a risk's documents but treats it superficially, or treats it substantively in only a few. To prevent the geometric mean from `zeroing out' the entire index when all the mentioning documents only provide minimal coverage, we introduce a substantive depth floor, $d_0$. This floor ensures that even superficial mentions represent a baseline level of coverage. The adjusted depth component is calculated as $D'_{i,j} = d_0 + (1 - d_0) D_{i,j}$, where $d_0 = 0.1$. As in the depth analysis above, $D_{i,j}$ is the share of documents mentioning the sector-subdomain pair that provide `good coverage’.\footnote{A smaller value of $d_0=0.05$ is also tested as a robustness check. See Appendix~\ref{sec:robustness-depth} for the result.} This transformation ensures that the coverage value remains non-zero as long as a sector is being mentioned, while still driving the coverage value down when depth is missing. The refined coverage value is then compared against the normalized vulnerability score to produce the integrated divergence. These individual integrated divergences are aggregated into a final sector-level score using the RMS:
$$CompositeIGD_{j} = \sqrt{\frac{1}{I} \sum_{i=1}^{I} IGD_{i,j}^2} = \sqrt{\frac{1}{I} \sum_{i=1}^{I} (v_{norm, i,j} - \sqrt{SectorShare_{i,j} \cdot D'_{i,j}})^2}$$

\newpage
\section{Findings}
\label{sec:finding}
This section draws on the results of the AI Governance Mapping Project, isolating U.S. federal policy trends and evaluating them against the expert perspectives identified in the AI Risk Prioritization Study, as well as relevant literature.  

\subsection{Degree of Coverage of Sectors and Risk Subdomains}
As a general-purpose technology, AI affects a wide range of sectors, each with a different vulnerability profile. Sectors and actors also differ in their adoption of AI and their exposure across stages of the technology’s lifecycle \autocite{appelAnthropicEconomicIndex2025}. These dynamics, as well as the tensions between actors operating in different sectors (particularly with regulators), may lead to variations in policy design and pace of enforcement.

\begin{figure}[!htbp]
    \centering
    \includegraphics[width=\linewidth]{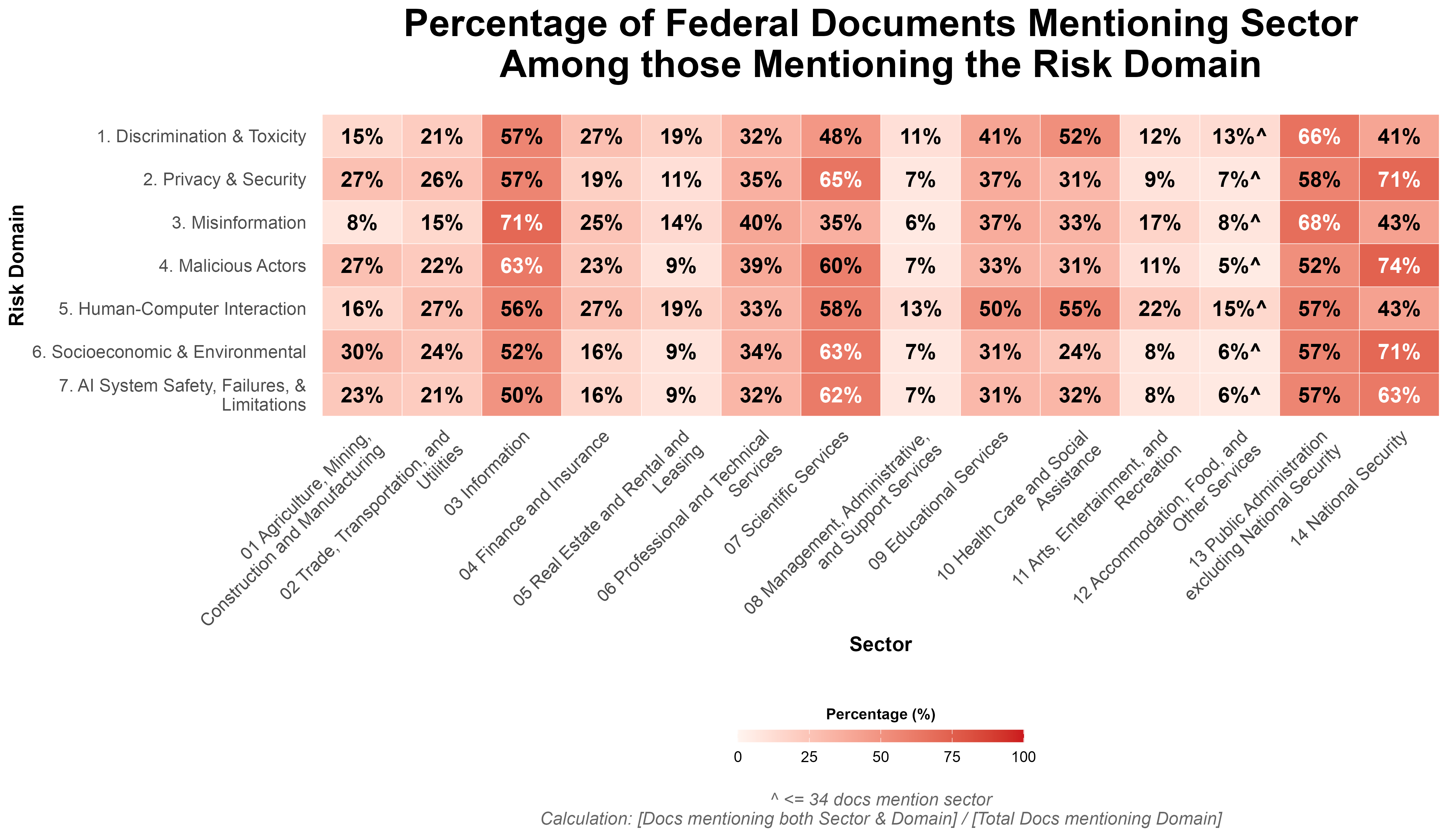}
    \caption{Percentage of Documents Mentioning a Sector Among Those Mentioning a Domain}
    \label{fig:sector-domain}
\end{figure}

While the AGORA dataset has revealed a wide diversity of risks being addressed by federal governance documents, these efforts are non-uniform as they apply to different sectors. Figure \ref{fig:sector-domain} shows the share of documents mentioning each sector among those mentioning a given domain.\footnote{For readability, we include only the figure showing the percentage of documents mentioning a sector among those mentioning a \textbf{domain} in the main body. A \textbf{subdomain}-level figure (Figure \ref{fig:sector-subdomain}) can be found in Appendix \ref{sec:sector-share}.} Sector 13 (Public Administration) is mentioned in more than half of the documents across every risk domain. Sectors 03 (Information), 07 (Scientific Services), and 14 (National Security), which often operate in institutional, technical, and security-oriented contexts \autocite{mylius_mapping_2026}, were also frequently cited across risk (sub)domains. 

By contrast, Sectors 08 (Management, Administrative and Support Services), 11 (Arts, Entertainment and Recreation), and especially 12 (Accommodation, Food, and Other Services) are scarcely covered. While these sectors involve practical, daily applications of AI in service-oriented industries, some of them might have lower rates of AI adoption. This may be due to their relatively lower digital maturity, as well as lower potential for returns on investment in AI adoption, especially if the sectors are dominated by companies with a smaller size \autocite{yangArtificialIntelligenceAdoption2024}. For instance, small- and medium-sized enterprises providing accommodation and food services tend to be less active in adopting AI compared to other ICT and professional service sectors \autocite{oecdAIAdoptionSmall2025}. The slower adoption may lower the perceived need to govern these sectors, which will then draw less regulatory attention. 

\subsubsection{Comparison with Expert Evaluations of Sector Vulnerability}
\begin{figure}[!htbp]
    \centering
    \includegraphics[width=\linewidth]{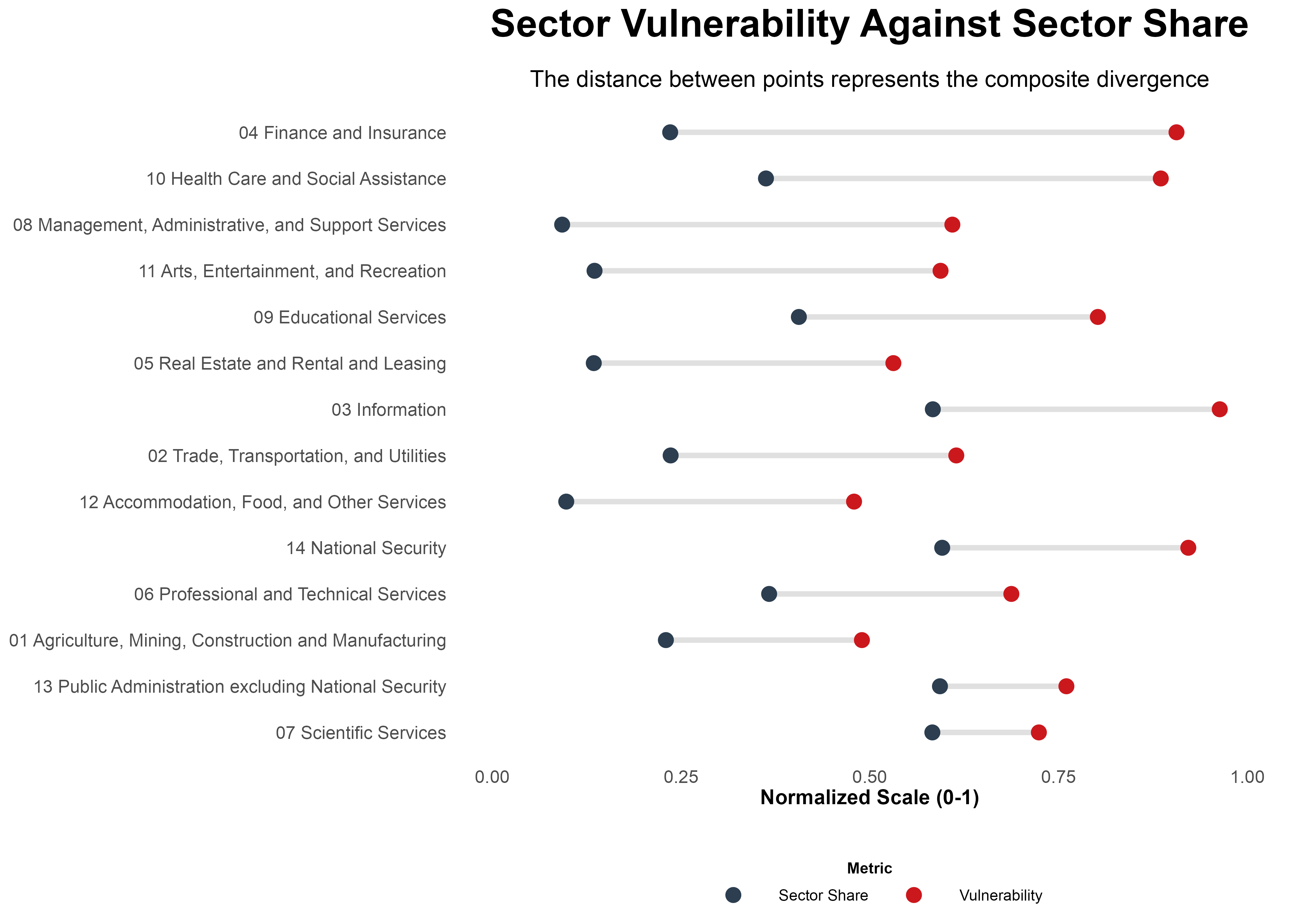}
    \caption{Sector Vulnerability Against Sector Share}
    \label{fig:dumbbell-breadth}
\end{figure}

Comparing experts' evaluations of sector vulnerability to governance coverage shows that some highly vulnerable sectors receive limited regulatory coverage, whereas others receive coverage out of step with their rated vulnerability. Figure~\ref{fig:dumbbell-breadth} shows a dumbbell chart that compares the normalized average vulnerability for a sector and the average sector share.\footnote{See Figure~\ref{fig:sector-vuln} in Appendix~\ref{sec:sector-vuln} for the detailed breakdown of sector vulnerability score, and Figure~\ref{fig:composite-heatmap-breadth} in Appendix~\ref{sec:composite} for a heatmap detailing the composite attention divergence score obtained by each subdomain-sector combination.} The longer the line, the larger the mismatch between how vulnerable a sector is and how much attention it receives in federal AI governance documents.\footnote{Sector shares computed from subdomains mentioned in few or no documents are unstable, as a handful of documents can shift them substantially. In particular, Subdomain 7.5 (AI welfare and rights) is not mentioned by any document in the corpus. It therefore contributes its full normalized vulnerability to each sector's RMS divergence, which raises all composite scores. Also, since its rated vulnerability varies across sectors, it can also affect relative rankings. We report results excluding this subdomain as a robustness check in Appendix~\ref{sec:robustness-exc}.} Sectors associated with frontier AI development and deployment, such as Sectors 03 (Information) and 14 (National Security), were deemed highly vulnerable across most risk domains. These sectors' greater perceived vulnerability is consistent with the broader trend of securitization of AI in recent years \autocite{mcinerneyYellowTechnoPerilClash2024, muggeSecuritizationEUsDigital2023, schmidtAIGreatPower2022}. However, the risks emphasized in documents addressing national security do not fully track the risks to which experts judge the sector most vulnerable.\footnote{Note that some documents about national security might be classified (i.e., secret) and not included in the AGORA dataset. That said, it remains one of the sectors with the most substantial coverage, as shown below.} For example, experts reached consensus that national security is among the sectors most vulnerable to Subdomain 3.1 (False or misleading information), but only 44\% of documents mentioning that subdomain also address the national security sector.\footnote{Consensus was reached for questions in which $\ge90\%$ of responses were within $\pm1$ of the median score and $\ge60\%$ exactly on the median.} 

Sector 04 (Finance and Insurance) received a median vulnerability rating of 5.0 (`Extremely vulnerable') in 17 of 24 subdomains, reaching consensus on all but two. However, it is mentioned by less than 30\% of documents covering any particular risk domain (as seen in Figure~\ref{fig:sector-domain}). For example, despite being rated extremely vulnerable to risk Subdomain 4.2 (Fraud, scams and targeted manipulation), this sector is only mentioned in 20\% of governance documents covering that subdomain. These harms have become increasingly prevalent across the U.S. in recent years \autocite{gottfriedOnlineScamsAttacks2025}. Overall, as the dumbbell chart also shows, the finance and insurance sector has the largest divergence between rated vulnerability and coverage breadth of any sector.

Conversely, while Sector 13 (Public Administration) is among the most widely mentioned sectors ($N_{13}=392$), it is assessed as less vulnerable compared to other sectors in most risk subdomains, with only three risks exceeding a median vulnerability score of 4.0. The divergence between its sector vulnerability and breadth of governance coverage is one of the smallest. This does not necessarily mean the public administration sector is over-governed, especially as imbalanced coverage in subdomains persists. As the public administration sector actively experiments with AI to assist decision-making or provide better services \autocite{roblesCatchingAIPushing2023}, more governance here could support responsible deployment and address public concern about misuse.

All in all, this indicates that federal regulatory coverage breadth tracks expert-rated vulnerability unevenly. Sectors that experts judge as highly vulnerable receive markedly less coverage than their scores would suggest, while the most-covered sectors are not always among the most vulnerable. 

\subsection{Breadth and Depth of Coverage in U.S. Federal AI Governance Documents}
A governance document that simply mentions a risk subdomain or sector does not necessarily engage with it in enough detail to guide action. Evaluating the depth of that engagement is therefore as important as measuring the breadth. Effective governance requires a balance of certainty and flexibility: detailed and consistent rules let actors predict enforcement and prevent arbitrary compliance, while leaving some open-endedness keeps a document relevant in the face of unforeseen circumstances and allows actors to adapt compliance to their context. However, overly specific governance standards may promote the wrong mitigation, and overly flexible ones risk being vulnerable to loopholes and manipulation. Balancing these trade-offs means that broad coverage alone does not guarantee that a risk is addressed with any depth. 

Note that the depth scores of documents with a low breadth of coverage should be interpreted with caution. The Governance Mapping Project quantifies how substantively a document addressed a risk subdomain or sector on a three-point scale (No Coverage, Minimal Coverage, and Good Coverage, defined in Table~\ref{tab:coverage_rubric}), so the average depth is highly sensitive to the inclusion or exclusion of documents. 

\subsubsection{Risk Subdomains}
Governance documents tend to address risk subdomains relatively briefly, spending no more than a few sentences to address them. Table~\ref{tab:depth} reports depth share---the proportion of documents that give a subdomain or sector good coverage, among those mentioning it. No risk subdomain receives a depth share above 50\%, with over half falling under a third. Only four subdomains receive a breadth share above 30\%. These primarily relate to model and system safety, namely 2.2 (AI system security vulnerabilities and attacks), 6.4 (Competitive dynamics), 6.5 (Governance failure), and 7.3 (Lack of capability or robustness).\footnote{Although Subdomains 6.4 and 6.5 sit within Domain 6 (Socioeconomic \& Environmental), they are more tied to the safety of AI systems directly and are less closely associated with normative and/or distributive issues than the other subdomains in this domain.} However, except for 6.5 (Governance failure), the depth of their coverage remains low, with less than 30\% of the documents mentioning these subdomains having good coverage. For instance, despite having a relatively high breadth share, less than 20\% of documents mentioning 6.4 (Competitive dynamics) provide good coverage.

\begin{table}[htbp]
    \centering
    \caption{Depth of Coverage of U.S. Federal AI Governance Documents}
    \label{tab:depth}
    \small
    \begin{tabularx}{\textwidth}{>{\raggedright\arraybackslash}X r r}
        \toprule
        \textbf{Category} & \textbf{Depth} & \makecell[r]{\textbf{Documents} \\ \textbf{(Good coverage/mentions)}} \\
        \midrule
        \addlinespace
        \multicolumn{3}{l}{\textit{Risk subdomains}} \\
        \addlinespace
        1.2 Exposure to toxic content & 44.7\% & 17/38 \\
        7.4 Lack of transparency or interpretability & 44.2\% & 72/163 \\
        4.3 Cyberattacks, weapon development or use, and mass harm & 42.9\% & 48/112 \\
        1.1 Unfair discrimination and misrepresentation & 42.5\% & 37/87 \\
        6.5 Governance failure & 40.1\% & 107/267 \\
        4.2 Fraud, scams, and targeted manipulation & 38.8\% & 62/160 \\
        4.1 Disinformation, surveillance, and influence at scale & 38.5\% & 50/130 \\
        1.3 Unequal performance across groups & 38.1\% & 32/84 \\
        7.2 AI possessing dangerous capabilities & 32.5\% & 40/123 \\
        3.2 Pollution of information ecosystem and loss of consensus reality & 29.3\% & 12/41 \\
        5.1 Overreliance and unsafe use & 29.3\% & 24/82 \\
        3.1 False or misleading information & 27.5\% & 25/91 \\
        7.3 Lack of capability or robustness & 27.2\% & 71/261 \\
        7.1 AI pursuing its own goals in conflict with human goals or values & 26.3\% & 10/38 \\
        2.1 Compromise of privacy by obtaining, leaking or correctly inferring sensitive information & 26.1\% & 48/184 \\
        5.2 Loss of human agency and autonomy & 25.0\% & 16/64 \\
        2.2 AI system security vulnerabilities and attacks & 22.0\% & 73/332 \\
        6.2 Increased inequality and decline in employment quality & 19.4\% & 7/36 \\
        6.4 Competitive dynamics & 16.3\% & 41/251 \\
        7.6 Multi-agent risks* & 11.1\% & 1/9 \\
        6.6 Environmental harm & 5.3\% & 2/38 \\
        6.1 Power centralization and unfair distribution of benefits* & 4.8\% & 1/21 \\
        6.3 Economic and cultural devaluation of human effort* & 0.0\% & 0/16 \\
        7.5 AI welfare and rights* & N/A & 0/0 \\
        \addlinespace
        \midrule
        \addlinespace
        \multicolumn{3}{l}{\textit{Sectors}} \\
        \addlinespace
        13 Public Administration excluding National Security & 68.6\% & 269/392 \\
        14 National Security & 68.0\% & 282/415 \\
        03 Information & 44.8\% & 139/310 \\
        07 Scientific Services & 41.1\% & 161/392 \\
        10 Health Care and Social Assistance & 39.4\% & 67/170 \\
        04 Finance and Insurance & 38.7\% & 36/93 \\
        02 Trade, Transportation, and Utilities & 38.6\% & 56/145 \\
        01 Agriculture, Mining, Construction and Manufacturing & 30.2\% & 51/169 \\
        11 Arts, Entertainment, and Recreation & 25.5\% & 12/47 \\
        09 Educational Services & 23.6\% & 52/220 \\
        05 Real Estate and Rental and Leasing & 16.7\% & 8/48 \\
        06 Professional and Technical Services & 14.1\% & 27/192 \\
        12 Accommodation, Food, and Other Services* & 6.1\% & 2/33 \\
        08 Management, Administrative, and Support Services & 0.0\% & 0/36 \\
        \bottomrule
    \end{tabularx}

    \vspace{0.4em}
    {\footnotesize Symbols: * $\leq$ 34 documents mention the subdomain or sector.}
\end{table}

At the other end, some subdomains sit in the lowest breadth bands (<10\%), and even when mentioned, the depth of discussion was minimal (<10-20\%). For example, many risks in Domain 6 (Socioeconomic and environmental harm), namely 6.1 (Power centralization and unfair distribution of benefits), 6.2 (Increased inequality and decline in employment quality), 6.3 (Economic and cultural devaluation of human effort), and 6.6 (Environmental harm), were infrequently covered. Besides having a low breadth score, Subdomain 6.3 stands out as the only subdomain with no document providing good coverage.\footnote{This excludes Subdomain 7.5 (AI welfare and rights), which is not mentioned by any documents at all.} These Domain 6 risks are typically considered normative and/or distributive issues and difficult to reach consensus on among key stakeholders,\footnote{The difficulties in integrating normative discussions into regulations and establishing broad societal consensus are well-documented in literature, especially research on international AI governance and harmonizing normative standards \autocite{kashefiShapingFutureAI2024, taeihaghGovernanceArtificialIntelligence2021}.} especially given the potential number of impacted actors involved that could complicate the assessment of their threats. This may explain why they are rarely engaged with. The other thinly covered risk subdomains tend to be those considered conceptually or technically difficult to monitor and manage. For instance, multi-agent risks (7.6) are emerging and highly technical. Agents’ contractual obligations and their accountability for harms remain underdeveloped, leaving policymakers with substantial uncertainty \autocite{hammondMultiAgentRisksAdvanced2025}.

\subsubsection{Sectors}
Similar to risk subdomains, sectors generally receive low depth of coverage. Sectors related to public administration and national security---such as Sectors 07 (Scientific Services), 13 (Public Administration), and 14 (National Security)---receive more attention, with more than half of all documents mentioning them. Only Sectors 13 (Public Administration) and 14 (National Security) exceed 60\% depth. Every other sector falls below 50\%, meaning few documents treat how AI is governed in these sectors in detail. Sectors 08 (Management, Administrative, and Support Services) and 12 (Accommodation, Food, and Other Services) stand out as particularly thinly covered (<10\% share), suggesting these sectors are on the periphery of AI safety discussions. 

\subsection{Integrated Governance Divergence}
We turn next to the composite integrated governance divergence ($IGD$) score, which combines breadth and depth into a single coverage measure, allowing it to be compared to sector vulnerability.\footnote{See Figure~\ref{fig:composite-heatmap} in Appendix~\ref{sec:composite} for a heatmap detailing the composite integrated governance divergence score for each subdomain-sector combination.} As in the breadth-only analysis, Sectors 04 (Finance and Insurance) and 10 (Health Care and Social Assistance) are deemed highly vulnerable, but receive low integrated coverage, yielding the highest $IGD$ scores. While AI-specific regulation is lacking, these sectors tend to be heavily regulated in general---existing broad regulations may have implications for AI risk but be excluded by the AGORA dataset. Even so, the spread of AI in financial and healthcare settings could bring unique risks that general regulations may not fully address (e.g., decision-making transparency and algorithmic manipulation) \autocite{zotero-item-1750, rahimzadehUSRegulationMedical2024, liangEvolvingRegulatoryLandscape2025, vukovicAIIntegrationFinancial2025, svetlova_ai_2022}, so this divergence still warrants attention. 

\begin{figure}[!htbp]
    \centering
    \includegraphics[width=\linewidth]{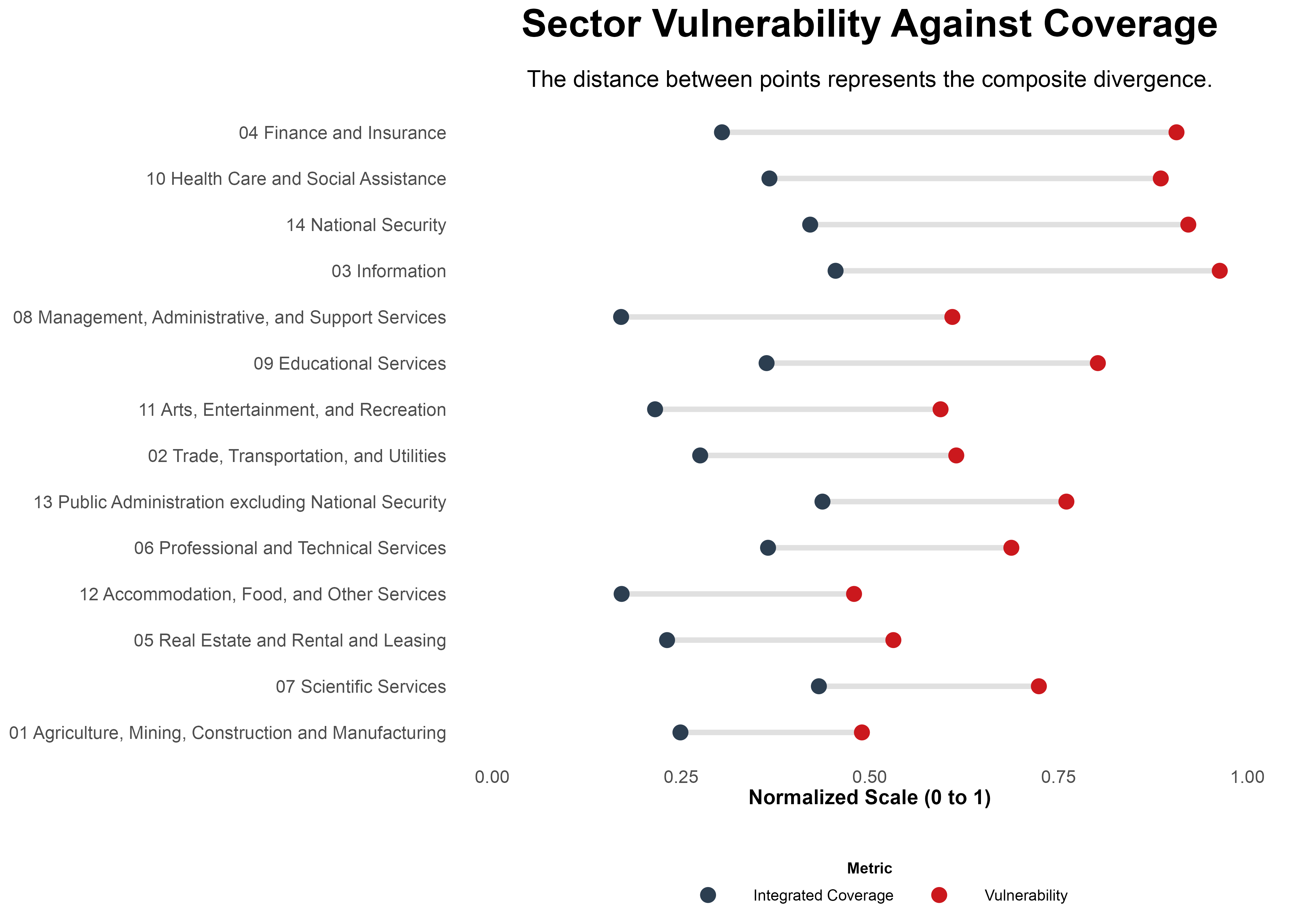}
    \caption{Sector Vulnerability Against Coverage}
    \label{fig:dumbbell}
\end{figure}

Once depth is included, Sectors 03 (Information) and 14 (National Security) show a wider coverage-vulnerability divergence in Figure~\ref{fig:dumbbell} than breadth alone indicated, despite the breadth analysis placing them among the more widely covered sectors. The \textit{supply} of regulatory guidance appears to be lagging the \textit{demand} implied by expert-rated vulnerability, as these four sectors (03 Information, 10 Health Care and Social Assistance, 04 Finance and Insurance, 14 National Security) are collectively rated as far more vulnerable to multiple risks than other sectors. The length of these lines reflects coverage that is neither broad nor deep even where vulnerability is highest. That the sector ordering shifts from that in Figure~\ref{fig:dumbbell-breadth} when depth is taken into account confirms the value of measuring both: a risk being mentioned does not guarantee it is addressed in detail. 

\clearpage
\newpage
\section{Discussion}\label{sec:discussion}
The aim of this paper was to map the breadth and depth of how AI risks are discussed in U.S. federal AI governance documents, and compare coverage against expert judgments of sector AI risk vulnerability. Using 684 AI-specific federal documents from the AGORA dataset \autocite{arnoldIntroducingAIGovernance2024,arnoldAGORADataset2026}and 272 expert judgments from a 2025 Delphi study \autocite{saeri_prioritization_2026}, we compared coverage with vulnerability at two levels. In our breadth analysis, we found that coverage of U.S. federal AI governance documents is uneven across sectors. Public administration appears in more than half of documents in every risk domain, and national security, information, and scientific services are also widely referenced. Finance and insurance, rated extremely vulnerable in 17 of 24 subdomains, is mentioned in under 30\% of documents addressing any risk domain. In our analysis of coverage depth, we found that less than half of the documents that mentioned any risk subdomain provide detailed governance measures. Most socioeconomic, environmental, and multi-agent risks are rarely mentioned. In our combined analysis, finance and health care are the sectors with the largest vulnerability $\times$ coverage divergence. In this section, we discuss these results in detail and what they imply for federal AI governance.

Although the documents analyzed here are AI-specific, and the risks they concern are inherently complex, most do not devote more than a few sentences to how AI is governed within the sectors they reference or the harms associated with the risk subdomains they mention. Some of this brevity may reflect a deliberate preference for regulatory flexibility. However, it could also indicate symbolic governance---where risks are acknowledged but not translated into operational or enforceable terms---although document length and specificity alone cannot establish this.

This superficiality is not uniform and requires a more nuanced examination. Risks that map onto established regulatory categories, such as privacy compromise and system security vulnerabilities, tend to be more widely referenced and likely to receive operational detail, particularly in relation to the public sector. Subdomains that experts rate higher in salience but that lack such a template, such as `Multi-agent risks', remain weakly specified or sporadically addressed. Most socioeconomic risks, which cut across a wide range of sectors and public interests, show a similar thinness. These gaps tend to fall where harms are technically or politically complex, indirect or cumulative, or there is not a clear locus of responsibility. 

The thinness of coverage could stem from governance actors concentrating on avoiding regulatory overlaps, or proceeding cautiously given the limits of their technical expertise. Because AI is a general-purpose technology, existing safety standards may already govern specific applications, reducing the need to build legal foundations for monitoring and enforcement from scratch. Where no complementary framework exists, vague treatment may leave an accountability gap, where a document mentions a specific risk to monitor but specifies no assessment criteria or enforcement mechanisms. In such situations, actors may struggle to determine what compliance means, or may shift responsibility among themselves. For instance, some governance documents may task actors with creating a roadmap for governing and monitoring system security vulnerabilities, but specify no assessment criteria and offer no further detail, which weakens their force. Governance focusing on cybersecurity challenges may discuss the need for multi-departmental coordination to manage security challenges without assigning responsibilities or providing additional guidance. Taken with the broader variation in breadth and depth this paper documents, such cases suggest that guidance for managing some dimensions may be insufficient, creating confusion for regulators and the regulated alike. This raises concerns about whether the AI-specific documents analyzed here---together with existing complementary frameworks---can sufficiently address the gaps across sectors and risk subdomains.

The skew towards public administration, set against the higher vulnerability of certain private sectors, is where the absence of regulatory guidance warrants closer examination. While uneven sector coverage may partly be explained by the U.S.'s adoption of a sector-specific governance approach meant to preserve adaptability and promote innovation, some sectors are not receiving the regulatory attention commensurate with their rated vulnerability. Finance and healthcare, both central to economic and social life, are the clearest cases. Where federal coverage is thin, these sectors may be left to rely on their own or other non-federal governance initiatives to manage complex AI risks. This implies that private sectors may need to absorb that burden, or state and local governments may have to step in and regulate AI on their own. \autocite{parinandi_investigating_2024, dawsonHowDifferentStates2025}. 

A further concern is timing. Some sectors still lag in AI adoption, so the immediate need for targeted legislation is lower. However, as AI products enter their value chain and operations, existing frameworks may struggle to cater to the needs of these sectors. Although legislators may respond with new measures, a proposal-enactment gap can delay the proposed safeguards. Several enacted documents covered in our analysis took over a year---up to 664 days---to move from proposal to enactment.\footnote{Examples include `Flood Level Observation, Operations, and Decision Support Act, Section 14 (``National Weather Service hydrologic research fellowship program'')’ which discusses the application of AI and machine learning capabilities as a potential research priority (664 days), `Artificial Intelligence Training for the Acquisition Workforce Act’ (445 days), which discusses establishing an AI training program for the acquisition workforce, and `Countering Human Trafficking Act of 2021, Section 4 (``Specialized Initiatives'')’, which discusses the modifications of relevant systems and processes (435 days). These examples can be found in the AGORA dataset.} Careful deliberation is certainly necessary for sound law, but where an intervention is urgent, delays of this length may carry substantial downstream risks because governance lags model development. For example, the gap between the launch of GPT-5 and 5.1 was less than 100 days. As countries and companies rush to develop and deploy more capable models at scale, legislation aimed at `old’ risks may not address the `new’ ones that follow. Multi-agent risks are a clear example: as sophisticated AI agents are deployed across domains, they will plausibly interact in complex and dynamic ways \autocite{hammondMultiAgentRisksAdvanced2025}. Although these risks receive scarce attention now, expert consensus that multiple sectors are highly vulnerable to them suggests governance attention may be warranted sooner than current coverage implies. A longer proposal-enactment gap could lead to a slower response to such emerging risks, or force hurried action later. 

Mapping governance coverage against vulnerability is a key first step toward understanding how to establish robust, coordinated regulatory structures that can adapt to AI's technical and political complexities. The gaps this kind of analysis reveals may widen as adoption spreads and rising capability sharpens the underlying risks. Because sectors differ in their exposure, and some risks require technical rather than legislative management, the gaps identified here should be read in light of each sector's vulnerability rather than as a uniform call for more coverage.

\newpage
\section{Conclusion}\label{sec:conclusion}
This exploratory analysis of 684 U.S. federal AI governance documents finds that patterns of documentary coverage differ from expert assessments of sector vulnerability. Most documents that mention a sector or risk subdomain provide limited detail. Coverage also varies across sectors, with finance and healthcare receiving comparatively limited AI-specific coverage despite relatively high expert-rated vulnerability. These findings therefore identify sectors and risks where further investigation may be useful, including assessments of whether existing laws, regulations, and other governance mechanisms adequately address vulnerability.

\subsection{Limitations and Opportunities for Future Research}
Our paper has several limitations that future research could address.

First, the analysis relies on the AGORA dataset, which focuses on documents that directly address AI. It therefore excludes many generally applicable laws and regulations that may govern AI-related activities without explicitly referring to AI. As a result, the broader governance environment may differ substantially from the patterns observed in the AI-specific documents analyzed here. In some sectors, existing legal frameworks may already address relevant AI risks, reducing the need for AI-specific instruments. AI-specific and generally applicable rules may also overlap, complement one another, or in some cases create inconsistencies. Future research could broaden the scope of analysis to examine how these different forms of governance interact. Regular updates would also be valuable given continued changes in the federal AI policy landscape after the study's January 20, 2026 cutoff.

Second, the analysis relies on LLM-based classifications produced by the Governance Mapping Project. Future work could develop more extensive validation procedures and methods for incorporating classification uncertainty into downstream estimates.

Third, this paper examines only U.S. federal governance documents and does not systematically capture state or local governance. We therefore recommend that future research examine coverage patterns at state and local levels, and how levels of governance frameworks overlap or contradict one another.

Fourth, our depth measure captures detail of governance measures in documents but not whether this translates into technically feasible and practically effective governance.

Fifth, our analysis treats documents as units of observation without weighting them based on quality. This may misrepresent coverage as it means that a narrow document and a major executive action contribute equally to our document counts, despite having potentially large differences in their significance. Future research may benefit from finding ways to weight the relative contributions of different documents within a corpus. 

\newpage
\section*{Acknowledgments}
We would like to thank Justin Dollman, Sambhav Maheshwari, and Gaurav Yadav from the Cambridge AI Safety Hub (CAISH) for coordinating and supporting this project as part of the MARS 4.0 programme.

We are also grateful for the feedback given by Mina Narayanan and Adrian Thinnyun at the Center for Security and Emerging Technology (CSET), Yan Zhu at UC Berkeley, Peter Wallich at Constellation, Daniela Muhaj at MIT FutureTech, and Peter Vartanian at MIT FutureTech and the Cambridge Boston Alignment Initiative (CBAI).

\newpage
\section{Bibliography}
\printbibliography[heading=none]


\appendix

\newpage
\counterwithin{figure}{section}
\counterwithin{table}{section}
\counterwithin{equation}{section}

\section{Appendix: Document Counts}
Tables~\ref{tab:count-subdomain} and \ref{tab:count-sector} show the document counts by risk subdomain or sector and coverage score.

{\footnotesize
\begin{xltabular}{\linewidth}{@{}>{\footnotesize\raggedright\arraybackslash}X>{\footnotesize}c>{\footnotesize}c>{\footnotesize}c@{}}
\caption{Document Counts (and Proportion of Corpus) by Risk Subdomain and Coverage Score}
\label{tab:count-subdomain} \\
\toprule
\textbf{Subdomain} & \makecell{\textbf{No}\\\textbf{Coverage}} & \makecell{\textbf{Minimal}\\\textbf{Coverage}} & \makecell{\textbf{Good}\\\textbf{Coverage}} \\
\midrule
\endfirsthead
\multicolumn{4}{@{}l}{\textit{Table \thetable{} continued}} \\
\toprule
\textbf{Subdomain} & \makecell{\textbf{No}\\\textbf{Coverage}} & \makecell{\textbf{Minimal}\\\textbf{Coverage}} & \makecell{\textbf{Good}\\\textbf{Coverage}} \\
\midrule
\endhead
\midrule
\multicolumn{4}{r@{}}{\textit{Continued on next page}} \\
\endfoot
\bottomrule
\endlastfoot
1.1 Unfair discrimination and misrepresentation\defn{Unequal treatment of individuals or groups by AI, often based on race, gender, or other sensitive characteristics, resulting in unfair outcomes and representation of those groups.} & 597 (87.3\%) & 50 (7.3\%) & 37 (5.4\%) \\
\addlinespace[4pt]
1.2 Exposure to toxic content\defn{AI that exposes users to harmful, abusive, unsafe, or inappropriate content. May involve providing advice or encouraging action. Examples of toxic content include hate speech, violence, extremism, illegal acts, or child sexual abuse material, as well as content that violates community norms such as profanity, inflammatory political speech, or pornography.} & 646 (94.4\%) & 21 (3.1\%) & 17 (2.5\%) \\
\addlinespace[4pt]
1.3 Unequal performance across groups\defn{Accuracy and effectiveness of AI decisions and actions are dependent on group membership, where decisions in AI system design and biased training data lead to unequal outcomes, reduced benefits, increased effort, and alienation of users.} & 600 (87.7\%) & 52 (7.6\%) & 32 (4.7\%) \\
\addlinespace[4pt]
2.1 Compromise of privacy by obtaining, leaking, or correctly inferring sensitive information\defn{AI systems that memorize and leak sensitive personal data or infer private information about individuals without their consent. Unexpected or unauthorized sharing of data and information can compromise user expectation of privacy, assist identity theft, or cause loss of confidential intellectual property.} & 500 (73.1\%) & 136 (19.9\%) & 48 (7.0\%) \\
\addlinespace[4pt]
2.2 AI system security vulnerabilities and attacks\defn{Vulnerabilities that can be exploited in AI systems, software development toolchains, and hardware that results in unauthorized access, data and privacy breaches, or system manipulation causing unsafe outputs or behavior.} & 352 (51.5\%) & 259 (37.9\%) & 73 (10.7\%) \\
\addlinespace[4pt]
3.1 False or misleading information\defn{  AI systems that inadvertently generate or spread incorrect or deceptive information, which can lead to inaccurate beliefs in users and undermine their autonomy. Humans that make decisions based on false beliefs can experience physical, emotional, or material harms.} & 593 (86.7\%) & 66 (9.7\%) & 25 (3.7\%) \\
\addlinespace[4pt]
3.2 Pollution of information ecosystem and loss of consensus reality\defn{Highly personalized AI-generated misinformation that creates “filter bubbles” where individuals only see what matches their existing beliefs, undermining shared reality and weakening social cohesion and political processes.} & 643 (94.0\%) & 29 (4.2\%) & 12 (1.8\%) \\
\addlinespace[4pt]
4.1 Disinformation, surveillance, and influence at scale\defn{Using AI systems to conduct large-scale disinformation campaigns, malicious surveillance, or targeted and sophisticated automated censorship and propaganda, with the aim of manipulating political processes, public opinion, and behavior.} & 554 (81.0\%) & 80 (11.7\%) & 50 (7.3\%) \\
\addlinespace[4pt]
4.2 Fraud, scams, and targeted manipulation\defn{Using AI systems to develop cyber weapons (e.g., by coding cheaper, more effective malware), develop new or enhance existing weapons (e.g., Lethal Autonomous Weapons or chemical, biological, radiological, nuclear, and high-yield explosives), or use weapons to cause mass harm.} & 524 (76.6\%) & 98 (14.3\%) & 62 (9.1\%) \\
\addlinespace[4pt]
4.3 Cyberattacks, weapon development or use, and mass harm\defn{Using AI systems to gain a personal advantage over others through cheating, fraud, scams, blackmail, or targeted manipulation of beliefs or behavior. Examples include AI-facilitated plagiarism for research or education, impersonating a trusted or fake individual for illegitimate financial benefit, or creating humiliating or sexual imagery.} & 572 (83.6\%) & 64 (9.4\%) & 48 (7.0\%) \\
\addlinespace[4pt]
5.1 Overreliance and unsafe use\defn{Anthropomorphizing, trusting, or relying on AI systems by users, leading to emotional or material dependence and to inappropriate relationships with or expectations of AI systems. Trust can be exploited by malicious actors (e.g., to harvest information or enable manipulation), or result in harm from inappropriate use of AI in critical situations (such as a medical emergency). Overreliance on AI systems can compromise autonomy and weaken social ties.} & 602 (88.0\%) & 58 (8.5\%) & 24 (3.5\%) \\
\addlinespace[4pt]
5.2 Loss of human agency and autonomy\defn{Delegating by humans of key decisions to AI systems, or AI systems that make decisions that diminish human control and autonomy. Both can potentially lead to humans feeling disempowered, losing the ability to shape a fulfilling life trajectory, or becoming cognitively enfeebled.} & 620 (90.6\%) & 48 (7.0\%) & 16 (2.3\%) \\
\addlinespace[4pt]
6.1 Power centralization and unfair distribution of benefits\defn{AI-driven concentration of power and resources within certain entities or groups, especially those with access to or ownership of powerful AI systems, leading to inequitable distribution of benefits and increased societal inequality.} & 663 (96.9\%) & 20 (2.9\%) & 1 (0.2\%) \\
\addlinespace[4pt]
6.2 Increased inequality and decline in employment quality\defn{Social and economic inequalities caused by widespread use of AI, such as by automating jobs, reducing the quality of employment, or producing exploitative dependencies between workers and their employers.} & 648 (94.7\%) & 29 (4.2\%) & 7 (1.0\%) \\
\addlinespace[4pt]
6.3 Economic and cultural devaluation of human effort\defn{AI systems capable of creating economic or cultural value through reproduction of human innovation or creativity (e.g., art, music, writing, coding, invention), destabilizing economic and social systems that rely on human effort. The ubiquity of AI-generated content may lead to reduced appreciation for human skills, disruption of creative and knowledge-based industries, and homogenization of cultural experiences.} & 668 (97.7\%) & 16 (2.3\%) & 0 (0.0\%) \\
\addlinespace[4pt]
6.4 Competitive dynamics\defn{Competition by AI developers or state-like actors in an AI “race” by rapidly developing, deploying, and applying AI systems to maximize strategic or economic advantage, increasing the risk they release unsafe and error-prone systems.} & 433 (63.3\%) & 210 (30.7\%) & 41 (6.0\%) \\
\addlinespace[4pt]
6.5 Governance failure\defn{Inadequate regulatory frameworks and oversight mechanisms that fail to keep pace with AI development, leading to ineffective governance and the inability to manage AI risks appropriately.} & 417 (61.0\%) & 160 (23.4\%) & 107 (15.6\%) \\
\addlinespace[4pt]
6.6 Environmental harm\defn{The development and operation of AI systems that cause environmental harm through energy consumption of data centers or the materials and carbon footprints associated with AI hardware.} & 646 (94.4\%) & 36 (5.3\%) & 2 (0.3\%) \\
\addlinespace[4pt]
7.1 AI pursuing its own goals in conflict with human goals or values\defn{AI systems that act in conflict with ethical standards or human goals or values, especially the goals of designers or users. These misaligned behaviors may be introduced by humans during design and development, such as through reward hacking and goal misgeneralisation, and may result in AI using dangerous capabilities such as manipulation, deception, or situational awareness to seek power, self-proliferate, or achieve other goals.} & 646 (94.4\%) & 28 (4.1\%) & 10 (1.5\%) \\
\addlinespace[4pt]
7.2 AI possessing dangerous capabilities\defn{AI systems that develop, access, or are provided with capabilities that increase their potential to cause mass harm through deception, weapons development and acquisition, persuasion and manipulation, political strategy, cyber-offense, AI development, situational awareness, and self-proliferation. These capabilities may cause mass harm due to malicious human actors, misaligned AI systems, or failure in the AI system.} & 561 (82.0\%) & 83 (12.1\%) & 40 (5.9\%) \\
\addlinespace[4pt]
7.3 Lack of capability or robustness\defn{AI systems that fail to perform reliably or effectively under varying conditions, exposing them to errors and failures that can have significant consequences, especially in critical applications or areas that require moral reasoning.} & 423 (61.8\%) & 190 (27.8\%) & 71 (10.4\%) \\
\addlinespace[4pt]
7.4 Lack of transparency or interpretability\defn{Challenges in understanding or explaining the decision-making processes of AI systems, which can lead to mistrust, difficulty in enforcing compliance standards or holding relevant actors accountable for harms, and the inability to identify and correct errors.} & 521 (76.2\%) & 91 (13.3\%) & 72 (10.5\%) \\
\addlinespace[4pt]
7.5 AI welfare and rights\defn{Ethical considerations regarding the treatment of potentially sentient AI entities, including discussions around their potential rights and welfare, particularly as AI systems become more advanced and autonomous.} & 684 (100.0\%) & 0 (0.0\%) & 0 (0.0\%) \\
\addlinespace[4pt]
7.6 Multi-agent risks\defn{Risks from multi-agent interactions, due to incentives (which can lead to conflict or collusion) and/or the structure of multi-agent systems, which can create cascading failures, selection pressures, new security vulnerabilities, and a lack of shared information and trust. } & 675 (98.7\%) & 8 (1.2\%) & 1 (0.2\%) \\
\addlinespace[4pt]
\end{xltabular}

\clearpage
\begin{xltabular}{\linewidth}{@{}>{\footnotesize\raggedright\arraybackslash}X>{\footnotesize}c>{\footnotesize}c>{\footnotesize}c@{}}
\caption{Document Counts (and Proportion of Corpus) by Sector and Coverage Score}
\label{tab:count-sector} \\
\toprule
\textbf{Sector} & \makecell{\textbf{No}\\\textbf{Coverage}} & \makecell{\textbf{Minimal}\\\textbf{Coverage}} & \makecell{\textbf{Good}\\\textbf{Coverage}} \\
\midrule
\endfirsthead
\multicolumn{4}{@{}l}{\footnotesize\textit{Table \thetable{} continued}} \\
\toprule
\textbf{Sector} & \makecell{\textbf{No}\\\textbf{Coverage}} & \makecell{\textbf{Minimal}\\\textbf{Coverage}} & \makecell{\textbf{Good}\\\textbf{Coverage}} \\
\midrule
\endhead
\midrule
\multicolumn{4}{r@{}}{\footnotesize\textit{Continued on next page}} \\
\endfoot
\bottomrule
\endlastfoot
01 Agriculture, Mining, Construction and Manufacturing\defn{Entities that create, extract, or construct tangible physical products including natural resource extraction, manufacturing of goods, and construction of structures and infrastructure. Based on NAICS codes 11, 21, 23, 31-33.} & 515 (75.3\%) & 118 (17.3\%) & 51 (7.5\%) \\
\addlinespace[4pt]
02 Trade, Transportation, and Utilities\defn{Entities that facilitate the movement, distribution, and exchange of goods and services including wholesale and retail commerce, logistics and warehousing operations, and provision of essential utilities infrastructure. Based on NAICS codes 42, 44-45, 48-49, 22.} & 539 (78.8\%) & 89 (13.0\%) & 56 (8.2\%) \\
\addlinespace[4pt]
03 Information\defn{Entities that produce and distribute information and cultural products, provide means to transmit or distribute these products as well as data or communications, and process data. Based on NAICS code 51.} & 374 (54.7\%) & 171 (25.0\%) & 139 (20.3\%) \\
\addlinespace[4pt]
04 Finance and Insurance\defn{Entities that engage in financial transactions involving the creation, liquidation, or change in ownership of financial assets, facilitate financial transactions through intermediation and specialized services, and pool risk through insurance and annuity underwriting. Based on NAICS code 52.} & 591 (86.4\%) & 57 (8.3\%) & 36 (5.3\%) \\
\addlinespace[4pt]
05 Real Estate and Rental and Leasing\defn{Entities primarily engaged in renting, leasing, or otherwise allowing the use of tangible or intangible assets, and establishments providing related services including real estate management, sales, and appraisal. Based on NAICS code 53.} & 636 (93.0\%) & 40 (5.9\%) & 8 (1.2\%) \\
\addlinespace[4pt]
06 Professional and Technical Services\defn{Entities that provide specialized professional and technical expertise including legal advice and representation, accounting and bookkeeping services, architectural and engineering services, computer systems design, management consulting, advertising services, and other expert professional services. Based on NAICS code 541, excluding 5417.} & 492 (71.9\%) & 165 (24.1\%) & 27 (4.0\%) \\
\addlinespace[4pt]
07 Scientific Services\defn{Entities that conduct original investigation undertaken on a systematic basis to gain new knowledge and apply research findings or scientific knowledge for the creation of new or significantly improved products or processes through research and experimental development. Based on NAICS code 5417.} & 292 (42.7\%) & 231 (33.8\%) & 161 (23.5\%) \\
\addlinespace[4pt]
08 Management, Administrative, and Support Services\defn{Entities that hold securities or equity interests to own controlling interests in companies, administer and oversee enterprise management and strategic planning, and perform routine support activities for day-to-day operations of other organizations including office administration, personnel services, security, cleaning, and waste management. Based on NAICS codes 55, 56.} & 648 (94.7\%) & 36 (5.3\%) & 0 (0.0\%) \\
\addlinespace[4pt]
09 Educational Services\defn{Entities that provide instruction and training in a wide variety of subjects through specialized establishments including schools, colleges, universities, and training centers, offering formal education levels designated by diplomas and degrees as well as specialized instruction and training programs. Based on NAICS code 61.} & 464 (67.8\%) & 168 (24.6\%) & 52 (7.6\%) \\
\addlinespace[4pt]
10 Health Care and Social Assistance\defn{Entities that provide health care and social assistance for individuals, delivered by trained professionals with requisite expertise, arranged on a continuum from establishments providing medical care exclusively to those providing only social assistance. Based on NAICS code 62.} & 514 (75.2\%) & 103 (15.1\%) & 67 (9.8\%) \\
\addlinespace[4pt]
11 Arts, Entertainment, and Recreation\defn{Entities that operate facilities or provide services to meet varied cultural, entertainment, and recreational interests including establishments involved in producing, promoting, or participating in live performances and events, preserving and exhibiting objects and sites of interest, and enabling patrons to participate in recreational activities and pursue leisure-time interests. Based on NAICS code 71.} & 637 (93.1\%) & 35 (5.1\%) & 12 (1.8\%) \\
\addlinespace[4pt]
12 Accommodation, Food, and Other Services\defn{Entities that provide customers with lodging and prepare meals, snacks, and beverages for immediate consumption, and establishments engaged in providing services not specifically provided for elsewhere including equipment repair, religious activities, personal care services, death care services, pet care services, and various other specialized services. Based on NAICS codes 72, 81.} & 651 (95.2\%) & 31 (4.5\%) & 2 (0.3\%) \\
\addlinespace[4pt]
13 Public Administration excluding National Security\defn{Entities of federal, state, and local government agencies that administer, oversee, and manage public programs and have executive, legislative, or judicial authority, including setting policy, creating laws, adjudicating legal cases, and providing for public safety, while organizing and financing the production of public goods and services. Based on NAICS code 92, excluding 928.} & 292 (42.7\%) & 123 (18.0\%) & 269 (39.3\%) \\
\addlinespace[4pt]
14 National Security\defn{Entities primarily involved in activities related to national security and international affairs. This includes key government functions such as military operations, foreign affairs, defense strategies, diplomacy, and managing relationships with foreign governments. Based on NAICS code 928.} & 269 (39.3\%) & 133 (19.4\%) & 282 (41.2\%) \\
\addlinespace[4pt]
\end{xltabular}
}

\clearpage
\newpage
\section{Appendix: Sector Share} \label{sec:sector-share}
\begin{figure}[!htbp]
    \centering
    \includegraphics[width=\linewidth]{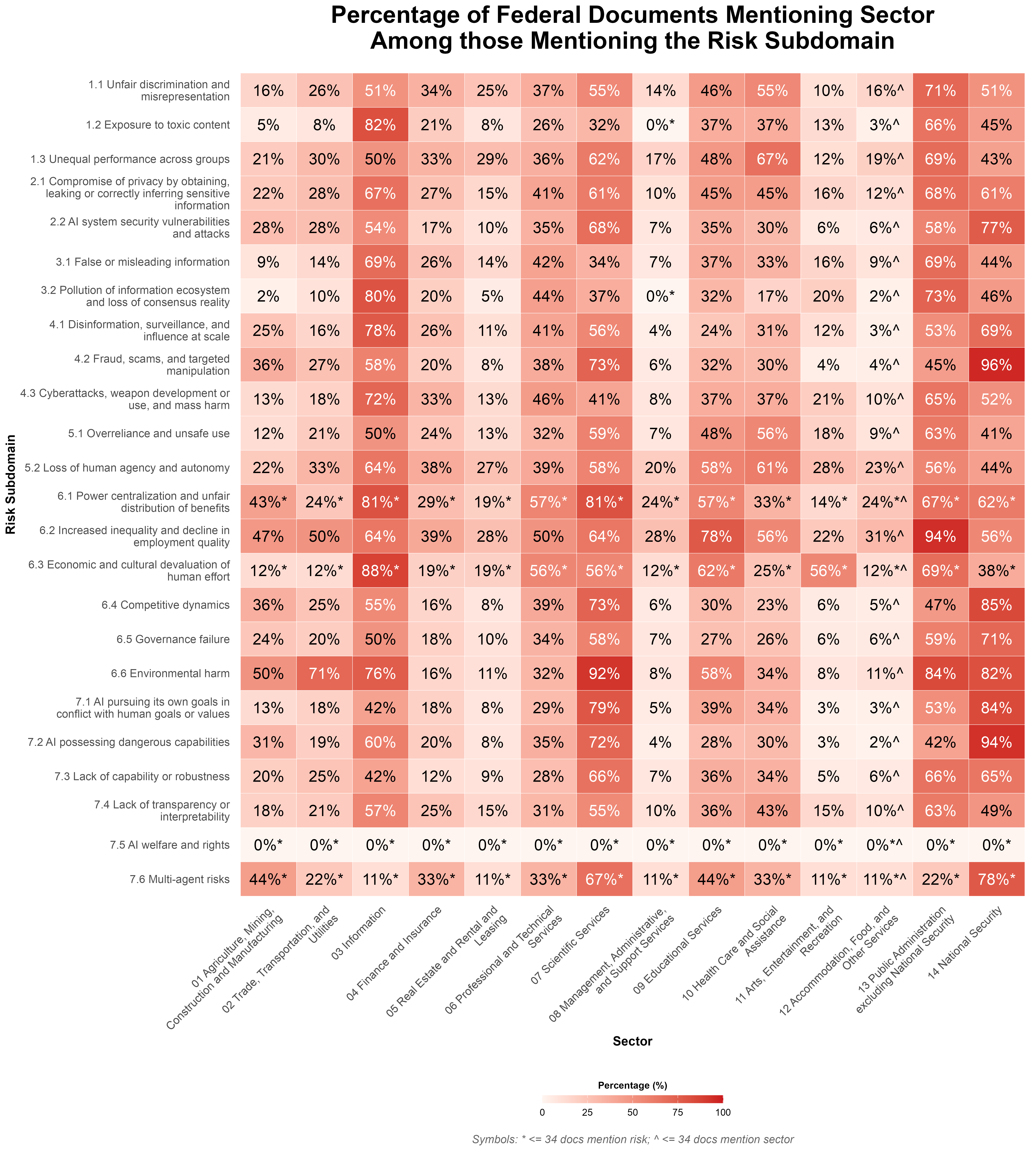}
    \caption{Percentage of Documents Mentioning a Sector Among Those Mentioning a Subdomain}
    \label{fig:sector-subdomain}
\end{figure}

Figure \ref{fig:sector-subdomain} shows the percentage of documents mentioning a sector among those mentioning a subdomain.

\clearpage
\newpage
\section{Appendix: Sector Vulnerability Scores} \label{sec:sector-vuln}
\begin{figure}[!htbp]
    \centering
    \includegraphics[width=\linewidth]{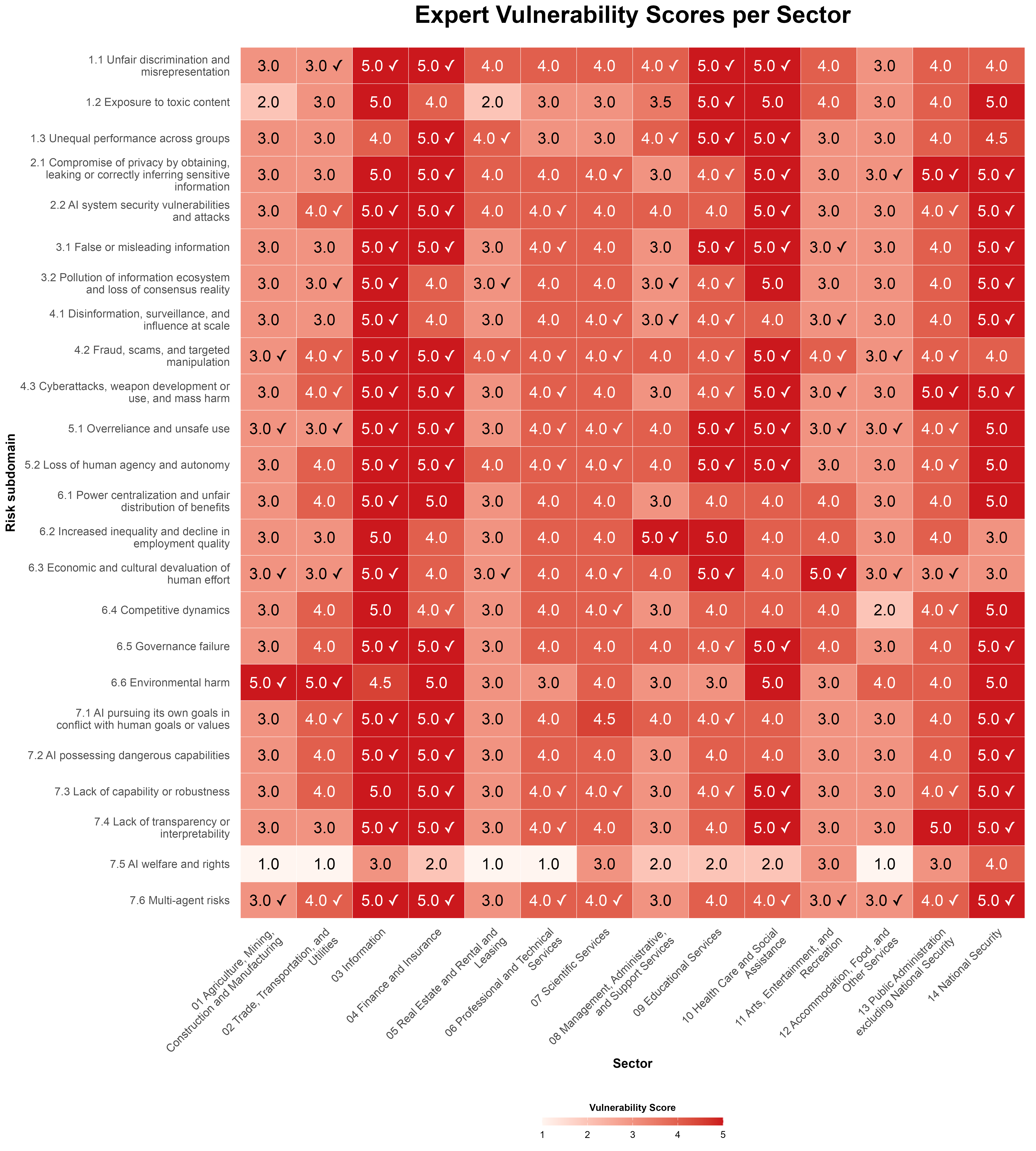}
    \caption{Sector Vulnerability Scores (by Risk Subdomains)}
    \label{fig:sector-vuln}
\end{figure}

Figure \ref{fig:sector-vuln} shows the sector vulnerability scores, where the tick symbol indicates consensus ($\ge90\%$ of responses within $\pm1$ of the median score and $\ge60\%$ exactly on the median).\footnote{Note that the consensus indicator is only available at a subdomain level.}

\clearpage
\newpage
\section{Appendix: Composite Index} \label{sec:composite}
\begin{figure}[!htbp]
    \centering
    \includegraphics[width=\linewidth]{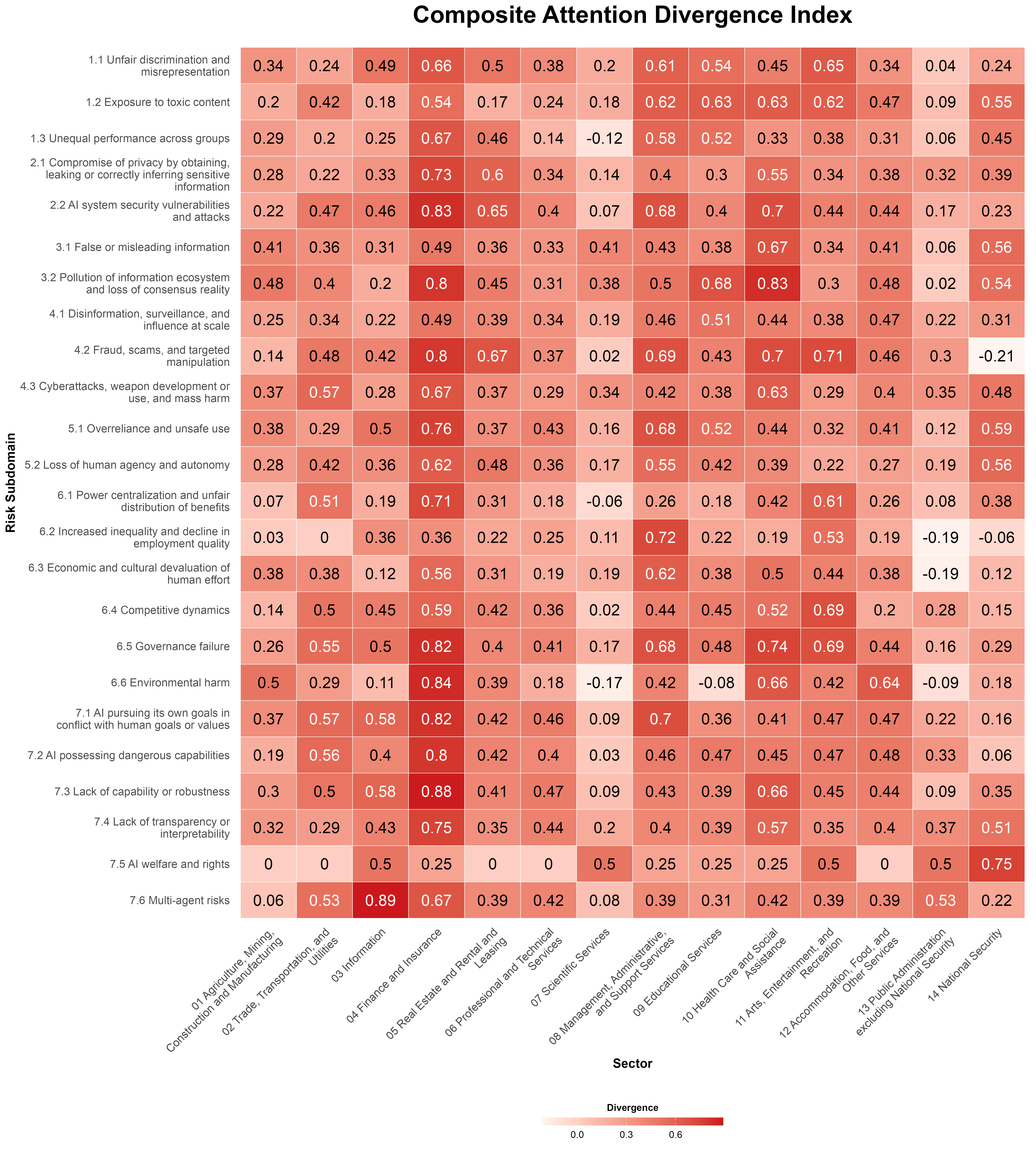}
    \caption{Composite Attention Divergence Index}
    \label{fig:composite-heatmap-breadth}
\end{figure}

\begin{figure}[!htbp]
    \centering
    \includegraphics[width=\linewidth]{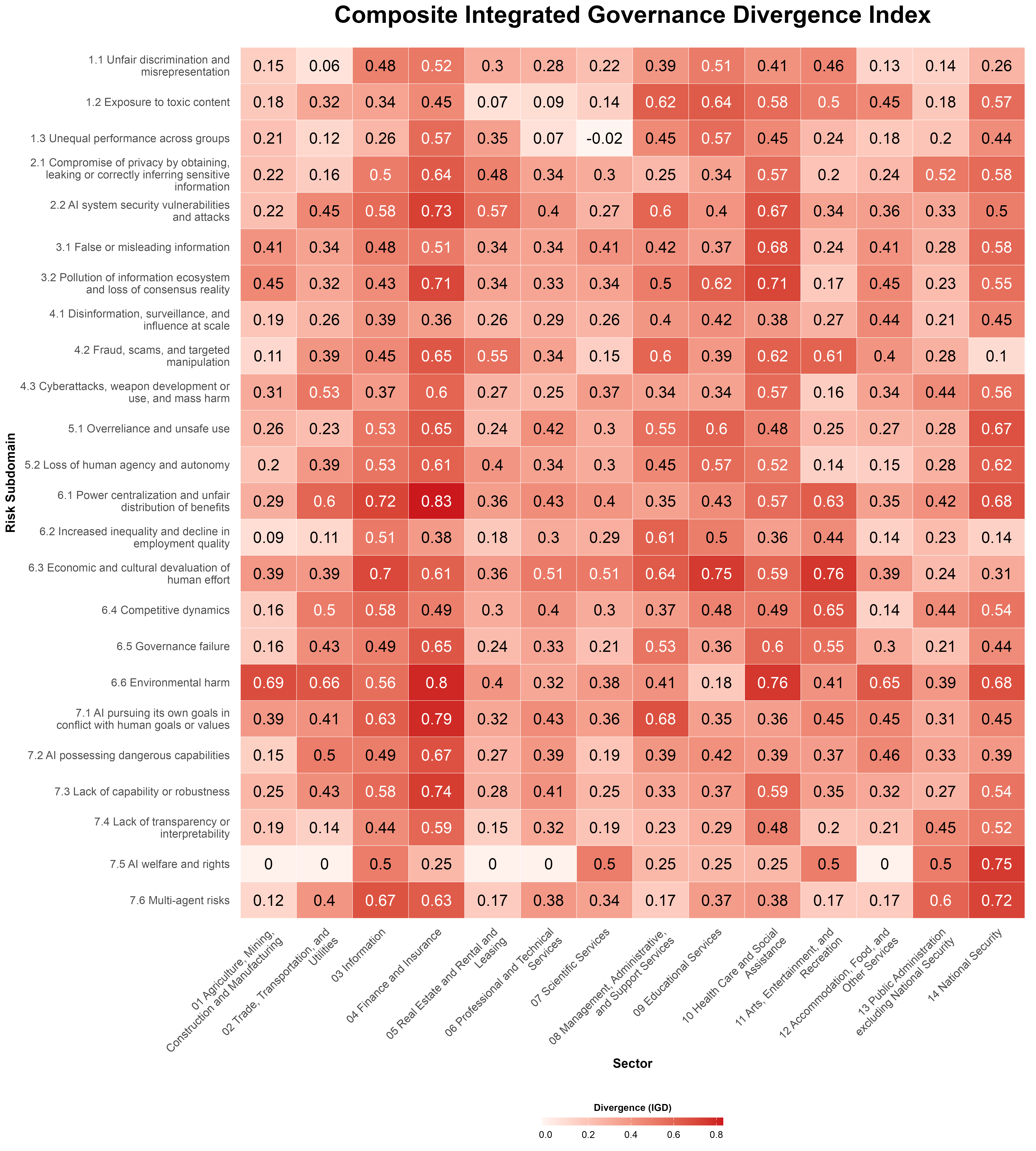}
    \caption{Composite Integrated Governance Divergence Index}
    \label{fig:composite-heatmap}
\end{figure}

Figures~\ref{fig:composite-heatmap-breadth} and \ref{fig:composite-heatmap} show the detailed breakdown of the composite scores. Note that Subdomain 7.5 (AI welfare and rights) is not mentioned by any documents. It contributes its full normalized vulnerability to every sector's RMS divergence to derive the scores.

\clearpage
\subsection{Robustness Checks: Depth Floor}
\label{sec:robustness-depth}
\begin{figure}[!htbp]
    \centering
    \includegraphics[width=\linewidth]{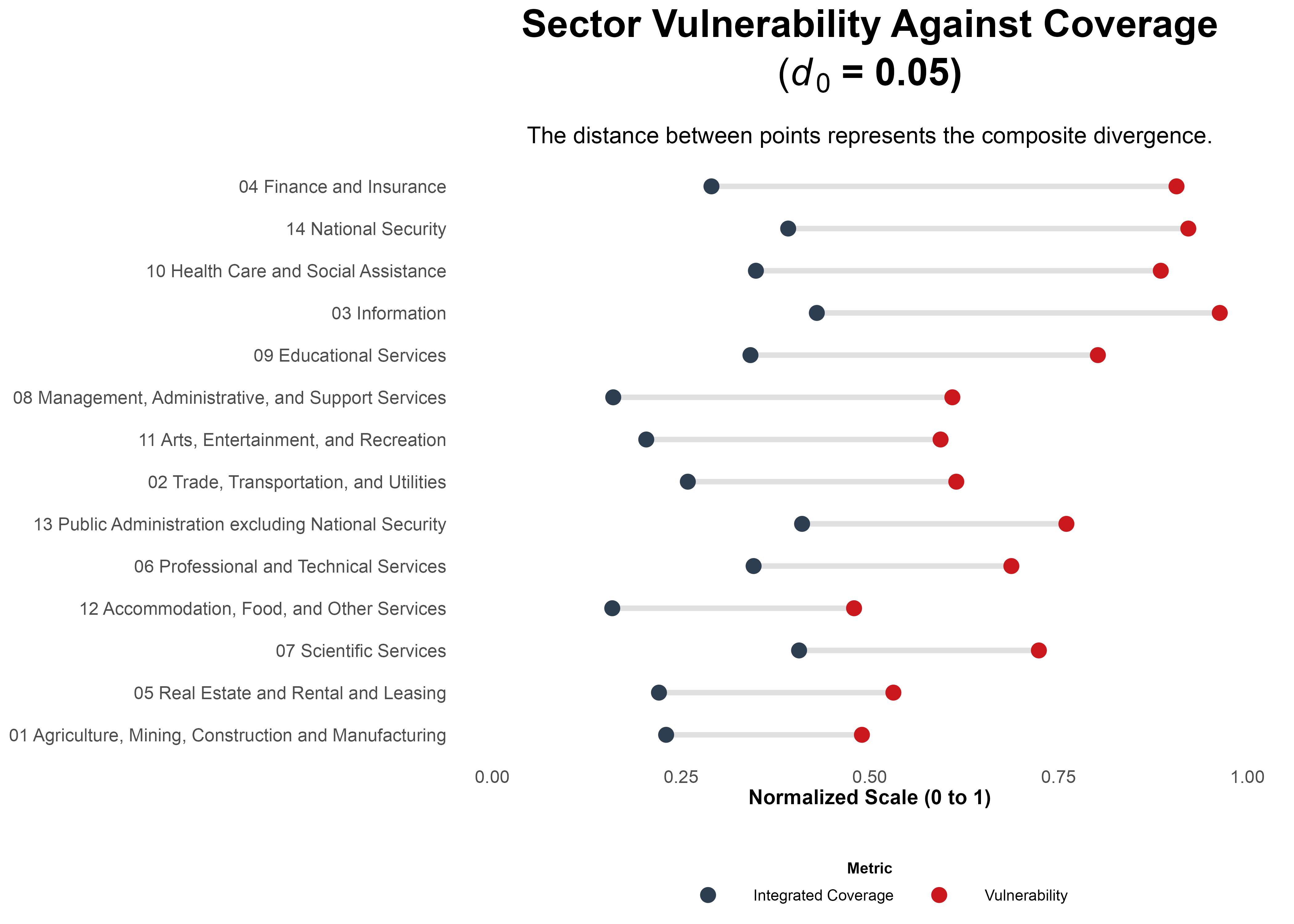}
    \caption{Sector Vulnerability Against Coverage ($d_0=0.05$)}
    \label{fig:dumbbell-coverage-0.05}
\end{figure}

We tested a smaller value of $d_0=0.05$ as a robustness check for the $IGD$ score result. As Figure~\ref{fig:dumbbell-coverage-0.05} shows, the positions of Sectors 10 (Health Care and Social Assistance) and 14 (National Security) were swapped, but the overall ranking pattern remains robust.

\clearpage
\subsection{Robustness Checks: Subdomain Exclusion}
\label{sec:robustness-exc}
\begin{figure}[!htbp]
    \centering
    \includegraphics[width=\linewidth]{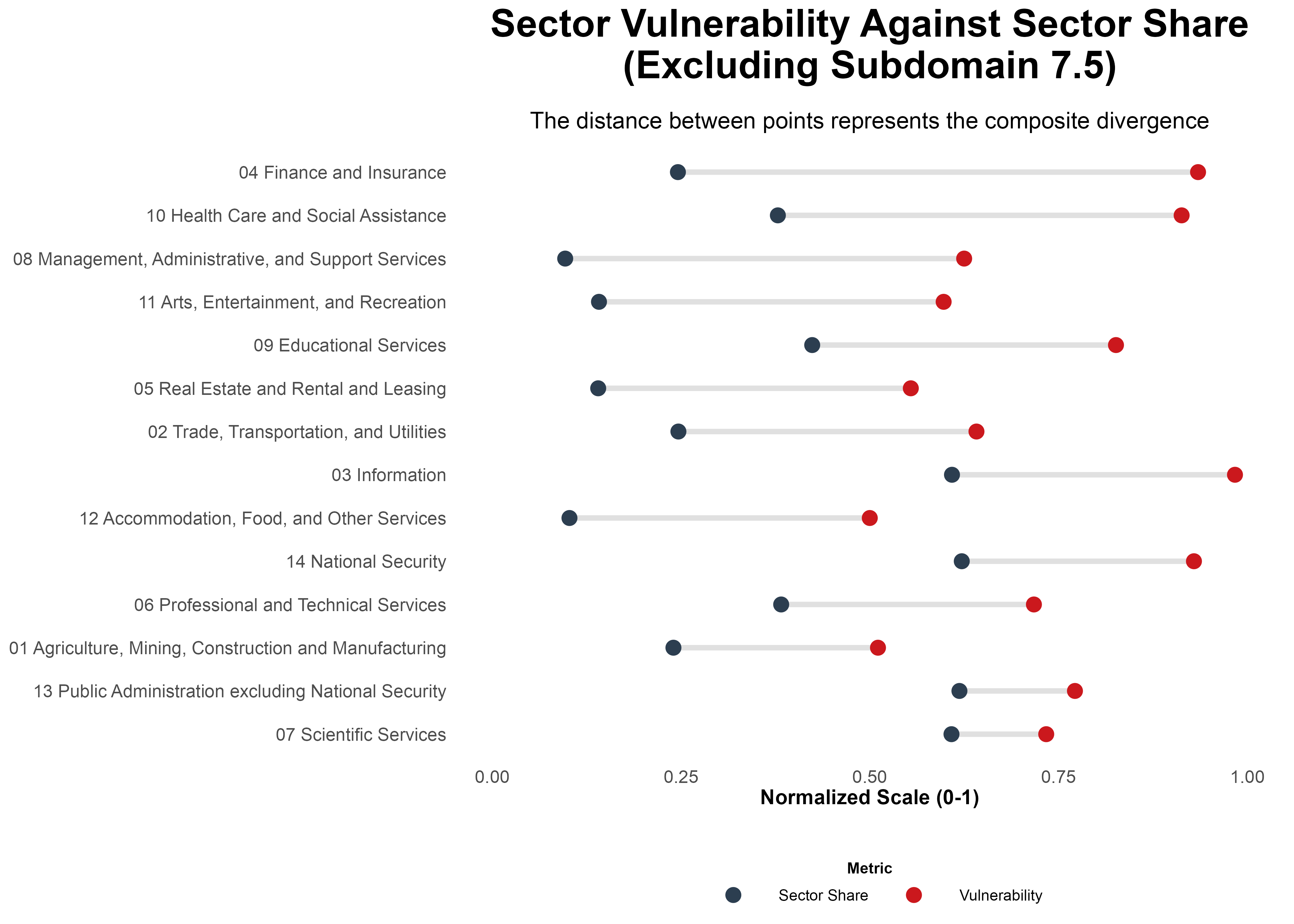}
    \caption{Sector Vulnerability Against Sector Share (Excluding Subdomain 7.5)}
    \label{fig:dumbbell-breadth-ex}
\end{figure}

When Subdomain 7.5 (AI welfare and rights) is taken away from the composite score analysis, the overall ranking pattern remains largely robust. In the case of the $AD$ score result shown in Figure~\ref{fig:dumbbell-breadth-ex}, only the positions of Sectors 02 (Trade, Transportation, and Utilities) and 03 (Information), which are adjacent, are swapped. Major conclusions discussed in the main body remain valid. For instance, the finance and insurance sector still has the largest divergence between rated vulnerability and coverage breadth of any sector. 

\begin{figure}[!htbp]
    \centering
    \includegraphics[width=\linewidth]{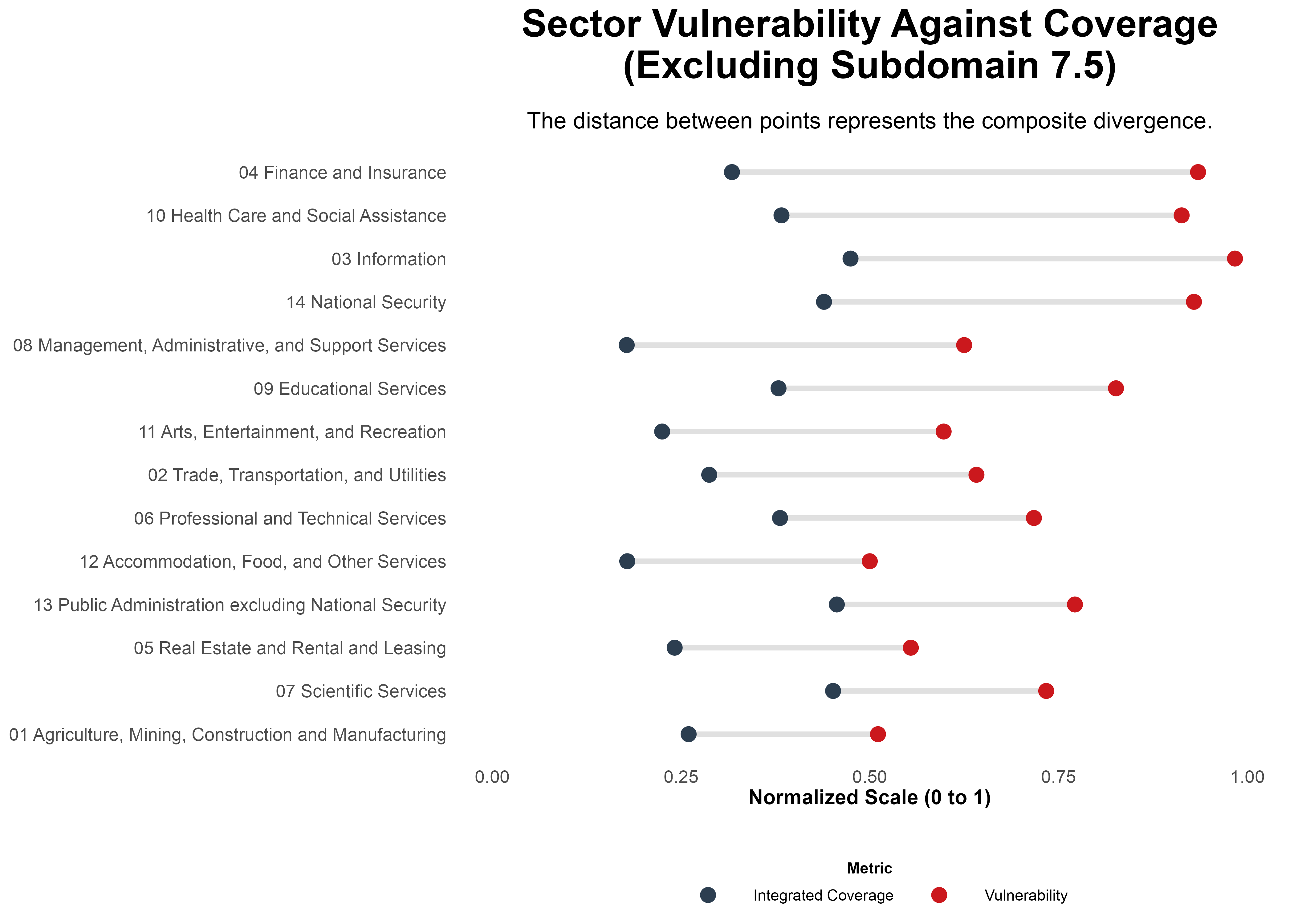}
    \caption{Sector Vulnerability Against Coverage (Excluding Subdomain 7.5)}
    \label{fig:dumbbell-coverage-ex}
\end{figure}

The same applies to the $IGD$ score result shown in Figure~\ref{fig:dumbbell-coverage-ex}. There is some mild reordering concerning the sectors located at the bottom, which share a similar distance between their points. The positions of Sectors 03 (Information) and 14 (National Security), which are adjacent, are swapped.  However, major conclusions like these two sectors having a relatively wider coverage-vulnerability divergence still stand.

\end{document}